\documentclass[
twocolumn, 
eqsecnum, prd, superscriptaddress, nofootinbib]{revtex4-2}

\usepackage{amsmath,amssymb}
\usepackage{amsthm}
\usepackage{mathtools}
\usepackage{appendix}
\usepackage{comment}
\usepackage{graphicx}
\usepackage{grffile}
\usepackage{mathrsfs}
\usepackage{bm}
\usepackage{url}
\usepackage{ulem}
\usepackage[colorlinks=true
,urlcolor=DARKBLUE
,anchorcolor=DARKBLUE
,citecolor=DARKBLUE
,filecolor=DARKBLUE
,linkcolor=DARKBLUE
,menucolor=DARKBLUE
,linktocpage=true
,pdfproducer=medialab
,pdfa=true
]{hyperref}
\usepackage{orcidlink}
\usepackage{subcaption}
\usepackage{caption}
\usepackage{xcolor}
\usepackage{soul}
\usepackage{algorithm}
\usepackage{algorithmic}
\usepackage{natbib}
\usepackage{physics}
\usepackage{siunitx}
\usepackage{acronym}
\usepackage{xspace}
\usepackage{tabularx}

\newcommand{\ee}{\mathrm{e}}

\newcommand{\bae}[1]{\begin{align} #1 \end{align}}

\definecolor{MONZA}{HTML}{CF000F}
\definecolor{DARKBLUE}{HTML}{00008b}
\definecolor{DARKMAGENTA}{HTML}{8b008b}
\definecolor{DARKRED}{HTML}{8b0000}
\definecolor{DARKORANGE}{rgb}{1.0, 0.55, 0.0}

\begin{document}
\title{Multi-band observation of lensed gravitational waves\\as a probe of small-mass dark matter halos}

\date{\today}
\author{Katsuya T. Abe}
\email{ktabecosmology@gmail.com}
\affiliation{Center for Frontier Science, Chiba University, 1-33 Yayoi-cho, Inage-ku, Chiba 263-8522, Japan}
\author{Shun Arai\,\orcidlink{0000-0002-2527-3705}}
\email{shunarai@kmi.nagoya-u.ac.jp}
\affiliation{Kobayashi-Maskawa Institute for the Origin of Particles and the Universe (KMI), Nagoya University, Nagoya, 464-8602, Japan}
\author{Ryoto Inui}
\email{inui.ryoto.a3@s.mail.nagoya-u.ac.jp}
\affiliation{Department of Physics, 
Nagoya University, Furo-cho Chikusa-ku, Nagoya 464-8602, Japan}

\author{Takahiro S. Yamamoto\,\orcidlink{0000-0002-8181-924X}}
\email{yamamoto.s.takahiro@resceu.s.u-tokyo.ac.jp}
\affiliation{Research Center for the Early Universe (RESCEU), The University of Tokyo, 7-3-1 Hongo, Bunkyo-ku, Tokyo 113-0033, Japan}

\author{Hirotaka Yarimoto}
\email{yarimoto0417@chiba-u.jp}
\affiliation{Department of Physics, Graduate School of Science,
Chiba University, 1-33 Yayoicho, Inage, Chiba 263-8522, Japan}

\author{Shuichiro Yokoyama\,\orcidlink{0000-0002-1811-5477}}
\email{shu@kmi.nagoya-u.ac.jp}

\affiliation{Kobayashi-Maskawa Institute for the Origin of Particles and the Universe (KMI), Nagoya University, Nagoya, 464-8602, Japan}
\affiliation{Department of Physics, 
Nagoya University, Furo-cho Chikusa-ku, Nagoya 464-8602, Japan}
\affiliation{Kavli Institute for the Physics and Mathematics of the Universe (WPI), The University of Tokyo, Kashiwa, Chiba 277-8583, Japan}

\date{\today}
\begin{abstract} 

The gravitational lensing effect of gravitational waves (GWs) has been extensively discussed as a probe of small-mass dark matter halos, which can provide missing information about dark matter.
We propose a multi-band observation of lensed GWs from a compact binary to observe both geometrical optics (GO) and wave optics (WO) effects from the same source.
This method is expected to be advantageous in breaking parameter degeneracies between a GW source and a dark matter halo acting as a lens.
We assume DECIGO or B-DECIGO as a space-based detector observing the early inspiral phase, and the ET as a ground-based detector observing the merger phase.
We perform a Fisher analysis of multi-band detection for a source with masses $m_1 = 30 M_{\odot}, m_2 = 20 M_{\odot}$ at redshift $z = 1.5$, and a lens with mass $3 \times 10^{3} M_{\odot}$ at redshift $z = 1.0$. With this setup, the GO effect appears in the ET frequency band, and the WO effect in that of DECIGO.
For the halo density profile, we adopt Singular Isothermal Sphere, Cored Isothermal Sphere (CIS), and Navarro-Frenk-White models. We find that multi-band observation resolves parameter degeneracies and significantly reduces errors in estimated parameters.
For the CIS model, in particular, we show that,  by combining ET and DECIGO observations, the lens mass error improves by about 71 $\%$ and 58 $\%$ compared to ET and DECIGO alone, respectively. Similarly, the impact parameter error is reduced by about 65 $\%$ and 70 $\%$, and the core size error by 34 $\%$ and 68 $\%$, respectively. From these results, we conclude that the multi-band observation of GWs from compact binaries improves the estimation of the lens object properties by breaking the parameter degeneracy.

\end{abstract}
\maketitle
\acresetall

\acrodef{UCMH}{ultra compact mini halo}
\acrodef{PBH}{primordial black hole}
\acrodef{CMB}{cosmic microwave background}
\acrodef{WO}{wave optics}
\acrodef{GO}{geometrical optics}
\acrodef{GL}{gravitational lensing}
\acrodef{LIGO}{Laser Interferometer Gravitational-Wave Observatory}
\acrodef{LISA}{Laser Interferometer Space Antenna}
\acrodef{DECIGO}{DECi-hertz Interferometer Gravitational-wave Observatory}
\acrodef{BBO}{Big Bang Observer}
\acrodef{ET}{Einstein Telescope}
\acrodef{CE}{Cosmic Explorer}
\acrodef{GLOC}{Gravitational-Wave Lunar Observatory for Cosmology}
\acrodef{PDF}{probability density function}
\acrodef{EoM}{equation of motion}
\acrodef{EoS}{equation-of-state}
\acrodef{GW}{gravitational wave}
\acrodef{RD}{radiation-dominated}
\acrodef{DM}{dark matter}
\acrodef{PTA}{pulsar timing array}
\acrodef{TDI}{Time Delay Interferometry measurement }
\acrodef{IMS}{Interferometry Metrology System}
\acrodef{SGWB}{stochastic gravitational wave background}
\acrodef{SIGW}{scalar-induced gravitational wave}
\acrodef{SNR}{signal to noise ratio}
\acrodef{CIS}{Cored Isothermal Sphere}
\acrodef{SIS}{Singular Isothermal Sphere}
\acrodef{GSIS}{Generalized Isothermal Sphere}
\acrodef{NFW}{Navarro Frenk White }
\section{Introduction}
\label{sec: intro}

In modern cosmology, unveiling the nature of \ac{DM}, which governs the structure formation throughout cosmic history, is one of the key questions to pursue the origin of the Universe. It has rapidly developed in the past three decades as the standard model of cosmology i.e. $\Lambda$-Cold-Dark-Matter ($\Lambda$CDM) model has been established by multiple observations \cite{SupernovaSearchTeam:1998fmf, SupernovaCosmologyProject:1998vns,Planck:2015fie, Planck:2018vyg, Planck:2019nip, Pan-STARRS1:2017jku, DES:2017myr, eBOSS:2020yzd}. 
While CDM successfully reproduces the large-scale structure of the Universe, it is incompatible with the nature of DM inferred from small-scale structures, such as satellite galaxies in the Milky Way~\cite{Klypin:1999uc, Koposov:2007ni} and galaxy clusters, as suggested by X-ray observations~\cite{Boyarsky:2014jta, Bulbul:2014sua}.
These observational implications motivate DM models beyond CDM, and there are many theoretical models, e.g., self-interacting CDM~\cite{Spergel:1999mh}, warm DM~\cite{Colombi:1995ze}, very light scalar particles~\cite{Marsh:2015xka, OHare:2024nmr, Hui:2016ltb}. 
The halos composed of these DM have core-like density profiles and a smaller abundance at the low-mass scale compared with the CDM model.

As an astrophysical probe of such a dark component in the Universe,
\ac{GL} is a useful phenomenon in which wave propagation is deflected and magnified by a gravitational field of massive objects~\cite{Schneider:1992bmb}. 
\Ac{GL} of electromagnetic waves has been utilized in various contexts, for instance, studying dark matter halo properties~\cite{DES:2017nhh,Mahler:2022tgu}, a search for the matter distribution of the universe with weak lensing~\cite{Bartelmann:2010fz}, a test of gravity theory through the observation of bending light by the Sun~\cite{Shapiro:2004zz}, the measurement of the Hubble constant from the supernova observation~\cite{Kelly:2023mgv, Pascale:2024qjr}, and exotic objects search through the microlensing observation~\cite{Niikura:2017zjd}.

Recently, \ac{GL} of \acp{GW} has also attracted interest. The LIGO-Virgo-KAGRA Collaboration conducted searches for \ac{GL} events during their third observing run; no such events were detected, and upper limits on the lensing probability were consequently established~\cite{LIGOScientific:2021izm, LIGOScientific:2023bwz}.
The pipeline development is being actively pursued~\cite{Chan:2024qmb}. \ac{GL} of \acp{GW} is also useful for exploring dark matter properties~\cite{Jung:2017flg, Urrutia:2021qak,Oguri:2020ldf, Oguri:2022zpn}.
Interestingly, in the case of \acp{GW}, because their typical wavelength is longer than that of electromagnetic waves, the so-called \ac{WO} effect could be important for the \ac{GL} by dark matter halos.
Roughly, whether the \ac{WO} effect can be negligible or not is determined by the comparison between the wavelength of the \acp{GW} and the Schwarzschild radius of the lens object, when the relation between the lens mass $M_{\rm L}$ and the wavelength of \ac{GW} $\lambda$ is $M_{\rm L} \lesssim 10^8 M_{\odot}(\lambda/10^{-5} ~\rm{pc})$, the \ac{WO} effect is not negligible. 
In Refs.~\cite{Nakamura:1997sw, Nakamura:1999uwi, Takahashi:2003ix}, such \ac{WO} effect has been discussed for the \ac{GL} of \acp{GW} from a supermassive binary black hole merger, with a typical wavelength of \(\sim 10^{-5}~\mathrm{pc}\), lensed by dark matter halos with masses of \(M \sim 10^8 M_\odot\). It was shown that, in such a system, the \ac{WO} effect could be observable by future space-based \ac{GW} detectors. 

Recent studies have shown in further detail that lensed \ac{GW} observations can constrain the parameters of the density profile of dark matter halos.
For instance, Cheung \textit{et al.}~\cite{Cheung:2024ugg} studied the \ac{GL} of \acp{GW} from binary black holes in \ac{LIGO} detector and showed that the diffraction effect of the lens object with the mass below $10^3 M_\odot$ plays an important role for estimating the halo properties. In Ref.~\cite{Tambalo:2022wlm}, it has been shown that the parameters of the dark matter halo as a lens object, including the lens mass, the slope of the matter density, and core size, can be determined with the accuracy of $\sim 1/\rm SNR$, where the SNR is the signal-to-noise ratio of the GWs.

In addition to the parameters of the \ac{GW} source and lens object themselves, there is another parameter that has a strong influence on the lensed \ac{GW} signal:
the {\it impact parameter}, which determines the relative position of \ac{GW} source from the lens object.
In the previous works, on the accuracy of the parameter determination, in particular, 
there exists a degeneracy between the impact parameter and lens mass, because the phase of \acp{GW} depends on them through the particular combination in \ac{GO} regime. While these studies are based on the single frequency band observation of individual different sources, we can observe the same source in multiple frequency bands as discussed in Refs.~\cite{Sesana:2016ljz, Muttoni:2021veo}.

In this work, we examine an idea to resolve the degeneracy of the impact parameter and the lens model parameters, applying a multi-band observation of the same source. We choose the combination of GW detectors \ac{DECIGO}~\cite{ Kawamura:2011zz} and \ac{ET}~\cite{Punturo:2010zz}. We also consider B-DECIGO~\cite{Kawamura:2018esd}, which is the scientific pathfinder of \ac{DECIGO}. By making use of these combinations, 
we can observe the early-inspiral phase as well as the merger phase of the black hole binary with the mass of $O(10) M_\odot$~\cite{Grimm:2020ivq, Nakano:2021bbw}.
In addition, we also expect to be able to observe the \ac{GO} and \ac{WO} regimes, assuming that the lens mass is $O(10^3) M_\odot$, and it would complement
the information of the impact parameter and the parameters of the lens object.
We perform the Fisher analysis to estimate the statistical error in the parameter estimation, and discuss how much accuracy is improved by the multi-band observations, and conclude whether we can break the degeneracy between the model parameters.
As a density profile of a halo as a lens object,
we consider \ac{SIS}, \ac{CIS}, and \ac{NFW}.

This paper is organized as follows.
First, we will give the general formulation for the \ac{GL} of \acp{GW} in Sec.~\ref{sec: formulation}. Following, we will summarize our lens models in Sec.~\ref{sec: lens_model} and see our waveform model and the Fisher analysis in Sec.~\ref{sec: fisher}. Finally, we present the results of the Fisher forecast in the multi-band observation in Sec.~\ref{sec: fisher} and give the discussion and conclusion in Sec.~\ref{sec: conclusion}. We will work in the units with $c=1$. Throughout our analysis, we assume the $\Lambda$CDM model and the cosmological parameters are taken from the results of $Planck\ 18$ \cite{Planck:2018vyg}.

%%%%%%%%%%%
%Sec.1
%%%%%%%%%%%
\section{Formulation of the gravitational lensing}
\label{sec: formulation}
\begin{figure}[h] 
	\centering
	\includegraphics[width=9cm]{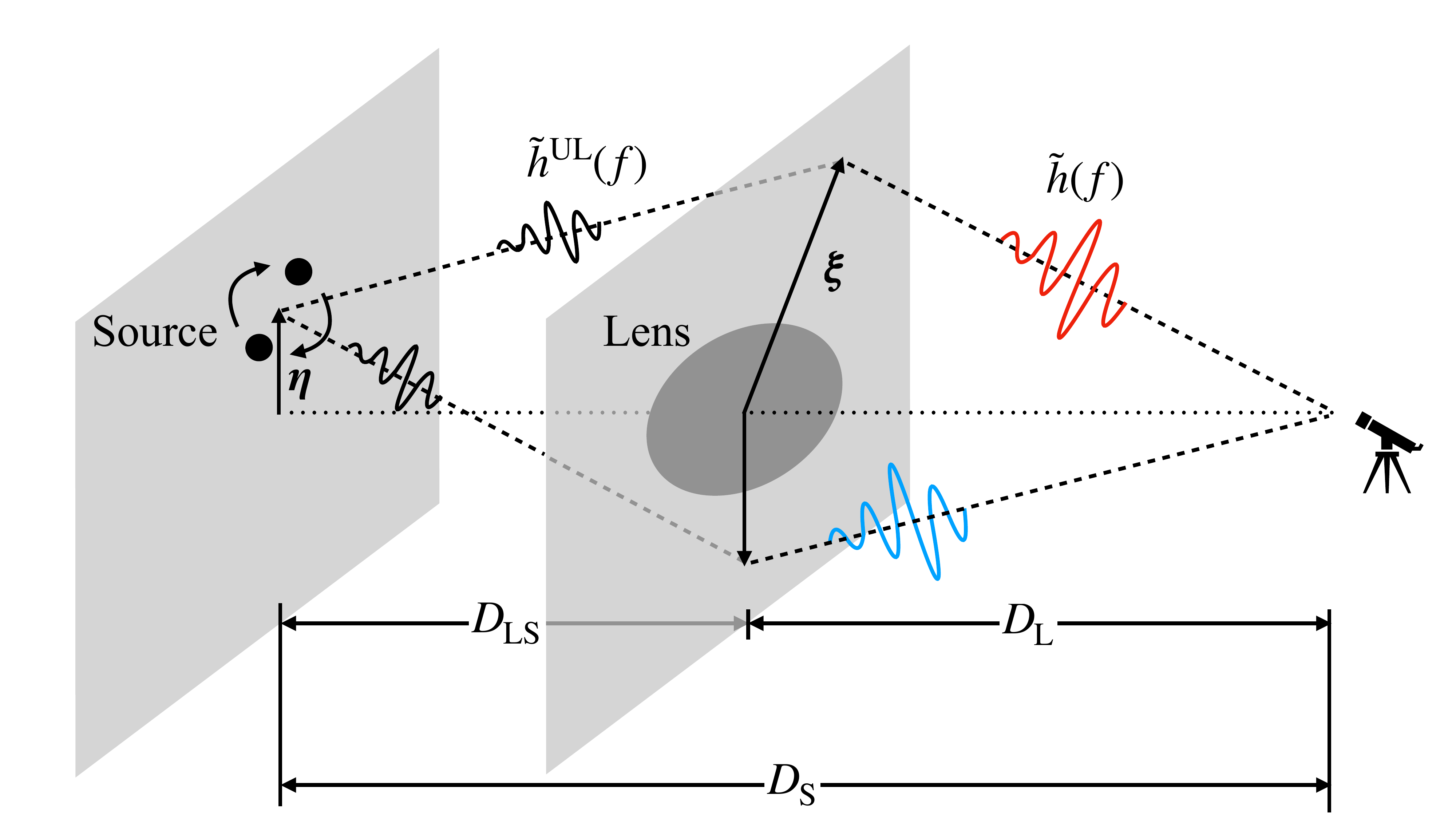}
	\caption{Schematic picture of the lens system. The two gray-shaded regions show the planes at $z_{\rm S}$ (left) and $z_{\rm L}$ (right) respectively.}
	\label{fig: lens_system}
\end{figure}

In this section, we provide the basic formulae for the \ac{GL} of \acp{GW}. 
Let us consider a lens system as shown in Fig.~\ref{fig: lens_system}. In this system, the propagation of \acp{GW} from the source to the observer can be investigated as the perturbation on the background spacetime given by a flat Friedmann-Lem\^aitre-Robertson-Walker (FLRW) metric with the Newton potential induced by the lens object:
\bae{\dd{s}^2 =g^B_{\mu\nu} \dd{x}^{\mu}\dd{x}^{\nu} = -a^2(1+2\Phi)\dd{\tau}^2 + a^2(1-2\Phi)\dd{\bf x}^2\,,}
which is shortly represented by $g^B_{\mu\nu}$ in the following explanation.  
Thus, let us consider the \ac{GW}, $h_{\mu\nu}$, as the linear perturbation on $g^B_{\mu\nu}$:
\bae{
g_{\mu\nu} = g^{B}_{\mu\nu} + a^2 h_{\mu\nu}~.}
Here, we adopt the weak gravitational field approximation, i.e, $\Phi({\bf x}) \ll 1$. 
The \ac{GW} source and the lens object are located at redshift $z_{\rm S}$ and $z_{\rm L}$, respectively. $D_{\rm S}$ and
$D_{\rm L}$ are angular diameter distances to the source and lens, respectively, and
$D_{\rm LS} = D_{\rm S} - \frac{1 + z_{\rm L}}{1 + z_{\rm S}} D_{\rm L}$
is the distance from the lens to the source. The two-dimensional vector $\bm \eta$ is a position vector pointing to the source on the source plane, and $\bm \xi$ denotes a two-dimensional position vector on the lens plane.
Suppose that the wavelength of \acp{GW} is much shorter than the scale of the background curvature, the propagation equation of \acp{GW} is approximately the same as that of a scalar function $h(\tau,{\bf x})$, namely $\partial_\mu(\sqrt{-g^B}g^{B\ \mu\nu}\partial_\nu h ) = 0$ with $h_{\mu\nu}(\tau,{\bf x}) = h(\tau,{\bf x})e_{\mu\nu}$. Here $e_{\mu\nu}$ denotes the polarization basis of \ac{GW} that satisfies the conditions of the parallel transport: $k^\alpha \partial_\alpha e_{\mu\nu} = 0$, $e^\mu_\mu = 0$, and $e^{\mu\nu} e_{\mu\nu} = 2$.   
Then, the field equation that governs the \ac{GL} in the Fourier domain is given as~\cite{Peters:1974gj}
\bae{
\left(\nabla^2  + \omega^2\right) \tilde{h} = 4\omega^2 \Phi \tilde{h} \,,
\label{eq: field eq}
}
where $\tilde{h}$ is the Fourier transformation in the frequency domain of the scalar function $h$ and $\omega$ denotes the angular frequency of the GW at the observer which is related to the GW frequency $f$ as $\omega = 2\pi f$. 
To characterize the lensing effect, it is useful to introduce the amplification factor defined by~\cite{Nakamura:1999uwi, Takahashi:2003ix}
\bae{
F \equiv \left. \frac{\tilde{h}}{\tilde{h}^{\rm UL}} \right|_\text{at observer}  ~,
}
where
$\tilde{h}^{\rm UL}$ is the unlensed wave solution obtained by solving Eq.~(\ref{eq: field eq}) with $\Phi = 0$.
For the distant observer ($D_{\rm LS}, D_{L}, D_{S} \gg |\bm \eta|, |\bm \xi|$), under the weak gravitational field approximation and the thin lens approximation,
the Born approximation can be applicable for solving Eq.~(\ref{eq: field eq}), and we can obtain an analytic expression for the amplification factor as~\cite{Nakamura:1999uwi, Takahashi:2003ix}%
\footnote{It has been known that the lensed wave solution can also be obtained by other methods, such as the Kirchhoﬀ diffraction integral~\cite{Schneider:1992bmb}, and the path integral approach~\cite{Nakamura:1999uwi}.}
\bae{
\label{eq: magnification_f}
F = F(w, \bm y) = \frac{w}{2\pi i}\int \dd^2{\bm x} \exp \left[i w T(\bm x,\bm y)\right]~,
}
with the dimensionless variables:
\bae{\bm{x} \equiv \frac{\bm{\xi}}{\xi_0}, ~\bm y \equiv \frac{D_{\rm L}}{\xi_0 D_{\rm S}} \bm{\eta},}
and
\bae{
w \equiv \frac{\xi_0^2}{D_{\rm eff}} \omega~, 
}
where $\xi_0$ is a dimensionful arbitrary length of the lens object, and $D_{\rm eff}$ is given by
\bae{D_{\rm eff} = \frac{D_{\rm L} D_{\rm LS}}{(1 + z_{\rm L}) D_{\rm S}}~. }
As we will see later, 
we consider the dark matter halo with a spherically symmetric density profile as the lens object.
To characterize the lens object, we introduce the effective redshifted lens mass $M_{{\rm L}z} \equiv (1 + z) M_\mathrm{L}$ to represent the scale of the lens mass. Then we adopt the normalization $\xi_0$ by
\bae{\xi_0 = \sqrt{4 G D_{\rm eff} M_{{\rm L}z}}\, ,\label{eq:xi0_to_M_Lz}
}
with the Newton's constant $G$. Note that $\xi_0$ is equivalent to the Einstein radius given a point-mass lens with the mass of $M_{{\rm L}z}$.
By using the effective redshifted lens mass, we have
\bae{w = 4 G M_{{\rm L}z} \omega~.}
Thus, $w$ represents the ratio between the Schwarzschild radius of the lens object and the wavelength of the \ac{GW}. In fact, the behavior of the integrand in Eq.~\eqref{eq: magnification_f} strongly depends on $w$; $w \gg 1$ corresponds to the \ac{GO} limit, and $w \ll 1$ corresponds to the \ac{WO} limit.
$T(\bm x, \bm y)$ in the integrand in Eq.~\eqref{eq: magnification_f} represents the time delay given a path of propagation, and it is given by 
\bae{
T(\bm x, \bm y) = \frac{1}{2}\abs{\bm x - \bm y}^2 - \psi(\bm x) - \phi_{\rm m}(\bm y)~,
}
where the first term on the right-hand side arises from the difference in the geometrical length between the lensed and the unlensed paths, and $\psi(\bm x)$ corresponds to the so-called deflection potential, which is the gravitational potential compressed onto the two-dimensional lens plane.
We set $\phi_{\rm m}(\bm y)$, 
which is the propagation time of the unlensed wave, so that the minimum value of the time delay is zero. 
Note that the derivation of the integral Eq.~\eqref{eq: magnification_f} does not depend on the property of the lens, while such properties are recast in $\xi_0$ and the deflection potential $\psi$ in the following way.

Given the three-dimensional density profile $\rho(\bm{x}, l)$ of a lens object composed of a thin single lens system, where $l$ denotes the coordinate of the line of sight, the deflection potential can be obtained by solving the Poisson equation in two dimensions as
\bae{
\label{eq: psi_kappa}
\nabla^2_x \psi(\bm x)  = 2\frac{\Sigma(\bm x)}{\Sigma_{\rm cr}},
}
where
$\Sigma(\bm x)$ is the surface mass density obtained by integrating the density profile over $l$ as
\bae{
\Sigma(\bm x) = \int^{\infty}_{-\infty} \dd{l} \left(\rho(\bm{x}, l) -\bar{\rho}(l) \right),
}
with the homogeneous background density $\bar{\rho}$,
and $\Sigma_{\rm cr}$ is the critical surface mass density defined as
\footnote{We often refer to the situation that the condition $\Sigma/\Sigma_{\rm cr}>1$ is satisfied as the strong lensing.}
\bae{
\Sigma_{\rm cr} \equiv \frac{1}{4 \pi G (1 + z_{\rm L}) D_{\rm eff}}~.  
}
Using the Green's function for the two-dimensional Laplacian is given by $G(\bm{x}, \bm{x}^{\prime}) = 1/2 \pi\log \abs{\bm{x} - \bm{x}^{\prime}}$,
the solution of Eq.~(\ref{eq: psi_kappa}) can be written as 
\bae{
\label{eq: deflection_potential}
\psi(\bm x) = \frac{1}{\pi}\int \dd^2{\bm{x}^{\prime}}\frac{\Sigma(\bm{x}^\prime)}{\Sigma_{\rm cr}}\log\abs{\bm{x} - \bm{x}^{\prime}}.
}
Thus, the signal of the lensed \acp{GW} depends on the parameters of the source through the waveform of $\tilde{h}^{\rm UL}$, and also those of the lens object through the amplification factor as a function of the frequency. In particular, the density profile of the dark matter halo as a lens object affects the amplification factor through the deflection potential. In the next section, we give the specific expression of the deflection potential for representative density profiles.

%%%%%%%%%%%%%%%%%%%%%%%%%%
Before closing this formulation section,
we would like to introduce one of the important observables in our setup.
As we will see later, in the Fisher analysis, we employ the black hole binary system with the primary mass $m_1 = 30 M_\odot$ and the secondary mass $m_2 = 20 M_\odot$ as the \ac{GW} source and the dark matter halo with the typical mass $M_{\rm L} = 3 \times 10^3 M_\odot$ as the lens object. In such a setup,
the ground-based detector such as \ac{ET} can observe the merger phase of the lensed \acp{GW} in the \ac{GO} regime, and it is expected to see the {\it time delay} of the \ac{GW} signal.
In the \ac{GO} limit, which corresponds to $w \gg 1$, the stationary points of the $T(\bm x, \bm y)$ dominate the integral in Eq.(\ref{eq: magnification_f}). The image positions can be determined by the lens equation $\nabla_{\bm x}T(\bm x, \bm y) = 0 $, i.e.
\bae{\label{eq: lens}
\bm{y} = \bm{x} - \nabla_x \psi(\bm x).
}
One can expand $T(\bm x, \bm y)$ around the $j$-th image position $\bm{x}_j$ up to the second order and can obtain the amplification factor in the \ac{GO} limit as
\bae{
F_{\rm GO}(w, \bm y) = \sum_j \abs{\mu_j}^{1/2} \exp \left[ i w T_j - i \pi n_j \right],
}
where $T_j = T(\bm{x}_j, \bm y)$, $\mu_j \equiv 1/\det \left|\partial_a \partial_b T \right|$ is the magnification of the $j$-th image given by the Hessian matrix $\partial_a \partial_b T$, and the phase shift $n_j = \{0, 1/2, 1\}$ corresponds to a minimum, saddle, and maximum points of the image respectively. The time delay between the $i$-th and $j$-th images is given by 
\bae{\label{eq time delay}
\Delta t_{\rm d} = \frac{\xi^2_0}{D_{\rm eff}}\left(T_i - T_j\right) = 4 G M_{\rm Lz} \Delta T\,,
}
where $\Delta T \equiv T_i - T_j$.

%%%%%%%%%%%%%%%%%%%%%%%

\section{Lense model}
\label{sec: lens_model}

In this section, we describe the density profiles of a dark matter halo used as a lens object in the analysis of the lensed \acp{GW},
and derive the corresponding deflection potentials appearing in the amplification factor. 
Here, for simplicity, we only consider spherically symmetric profiles.
Note that we do not take into account the \ac{GL} induced by the substructure in the dark matter halo or perturbers near the line of sight, such as subhalos,  focusing on the effects of a parent halo (see \cite{Poon:2024zxn} for such kind of studies). 
\subsection{Singular Isothermal Sphere (SIS)}
The \ac{SIS} is used for modelling a galaxy or a galaxy cluster with a constant velocity dispersion $v$. The density profile and the surface mass density are written as
\bae{
\rho(r) &= \frac{v^2}{2\pi G r^2}, \nonumber\\
\Sigma(x) &= \frac{v^2}{2 G \xi_0 x}.
}
Here and later in this section, $r = \sqrt{\xi_0^2 x^2 + l^2}$ denotes the radial distance from the lens. 
For simplicity, we assume $v^2$ as
\bae{v^2 = G \Sigma_{\rm cr}\xi_0\,,\label{eq:v2_SIS}}
so that $\xi_0$ is equivalent to the Einstein radius of the system.
and then $\Sigma ({\bm x})/\Sigma_{\rm cr} = 1/(2x)$.
The deflection potential takes the simple form as
\bae{
\psi(x) = x.
}
The analytical expression of the amplification factor for the SIS lens is given by~\cite{Matsunaga:2006uc} 
\bae{
F(w, y) &= \ee^{i w y^2/2} \sum^{\infty}_{n=0}\frac{\Gamma\left(1 + \frac{n}{2}\right)}{n!}\left(2 w \ee^{i3\pi/2}\right)^{n/2}\nonumber\\
&\times{}_1F_1\left(1 + \frac{n}{2}, 1; -\frac{i w y^2}{2}\right).
}
As can be seen in this expression, in the \ac{SIS} model, as a function of the frequency $f$, the amplification factor is basically controlled by two parameters $\xi_0$ (in $w$) and $y$. The frequency dependence of the amplification factor for $y = 0.3$ is displayed by the blue solid line in Fig.~\ref{fig: ampfactor_SIS_CIS_NFW}.
For the \ac{SIS} model, the amplification factor in the \ac{GO} regime is analytically given by
\bae{
F_{\rm GO}(w, y) = \begin{cases} 
   \abs{\mu_{+}}^{1/2} - i \abs{\mu_{-}}^{1/2} \exp\left(iw\Delta T\right) & \text{for } y < 1, \\
   \abs{\mu_{+}}^{1/2} & \text{for } y \geq 1,
\end{cases}
}
where $\mu_{\pm} = \pm1 + 1/y$ and $\Delta T =2y$. A single image and double images are created for $y \geq 1$ and $y < 1$, respectively.
The time delay given by Eq.~\eqref{eq time delay} is $\Delta t_{\rm d} \simeq 0.03\left(\frac{M_{\rm Lz}}{3\times10^3M_{\odot}}\right)\left(\frac{y}{0.3}\right)\;\rm s$.

\subsection{Cored Isothermal Sphere (CIS)}
The \ac{CIS} is the deformed model of the \ac{SIS} and it has the core radius $r_{\rm c}$~\cite{1994A&A...284..285K, Flores:1995dc, Treu:2010uj}.
The density profile and the surface mass density are written as
\bae{
\rho(r) &= \rho_0\frac{r_{\rm c}^2}{r^2 + r_{\rm c}^2}, \nonumber\\
\Sigma(x) &=  \frac{\pi \rho_0x_{\rm c}\xi_0}{\sqrt{(x/x_{\rm c})^2 + 1}}
\label{eq:xi_0_rE_SIS}}
where $\rho_0$ denotes the central density. We introduce $x_{\rm c} \equiv r_{\rm c}/\xi_0$. For simplicity, we assume $\rho_0$ as
\bae{\rho_0 = \frac{\Sigma_{\rm cr}}{2\pi\xi_0x^2_c}\,,\label{eq:rho0_CIS}}
so that $\xi_0$ is equivalent to the Einstein radius of the system.
The deflection potential can be written as
\bae{
\psi(x) = \sqrt{x^2 + x_{\rm c}^2} + x_{\rm c} \log\left(\frac{2x_{\rm c}}{\sqrt{x^2 + x_{\rm c}^2} + x_{\rm c}}\right).
}
To get the amplification factor, 
we have to solve Eq.(\ref{eq: magnification_f}) numerically.
In the \ac{CIS} model, in addition to $\xi_0$ and $y$, the amplification factor depends on the core radius $r_{\rm c}$.
The frequency dependence of the amplification factor for $y = 0.3,~x_{\rm c} = 0.3$ is exhibited by the orange-dashed line in Fig.~\ref{fig: ampfactor_SIS_CIS_NFW}.

\subsection{Navarro Frenk White (NFW) lens}
\begin{figure}[h] 
	\centering
	\includegraphics[width=9cm]{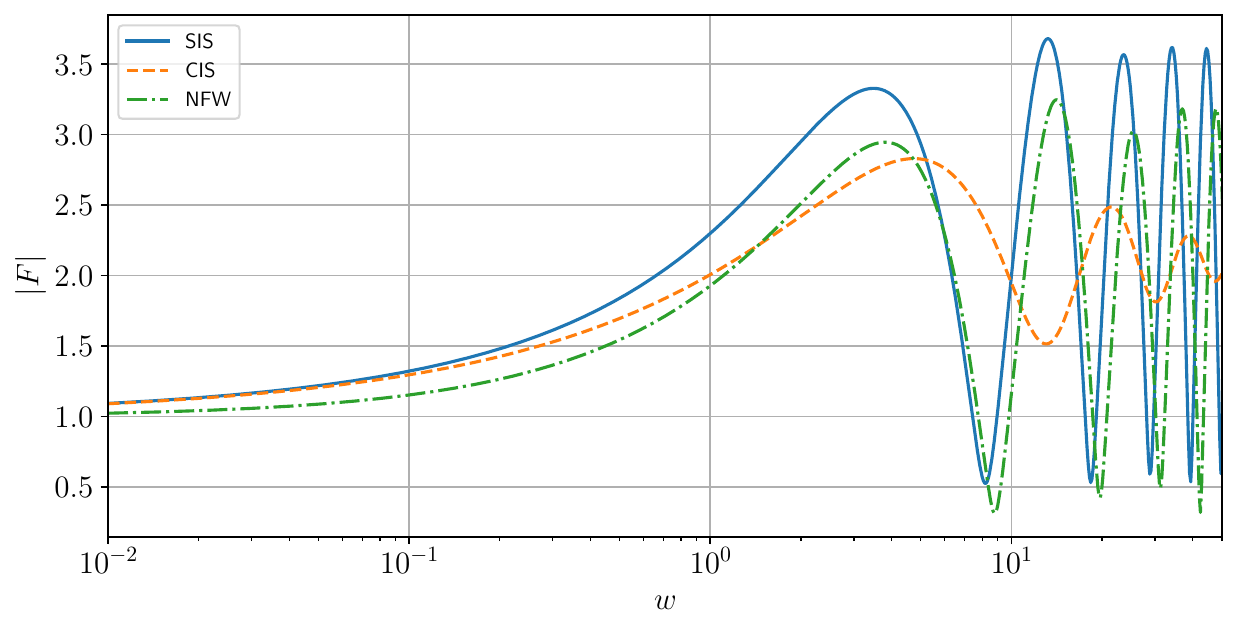}
	\caption{Amplification factors for the SIS, CIS, and NFW models. We set $y = 0.3$ for all lens models and $x_{\rm c} = 0.3$ for the CIS model, and  $x_{\rm s} = 0.3$ for the NFW model. The signals are numerically calculated by the public package $\rm GLoW$ ~\cite{Villarrubia-Rojo:2024xcj}}
	\label{fig: ampfactor_SIS_CIS_NFW}
\end{figure}

The \ac{NFW} profile is the common profile for the cold dark matter halo~\cite{Navarro:1995iw, Navarro:1996gj}.
The density profile and the surface mass density of the \ac{NFW} are given by
\bae{
\rho(r) = \frac{\rho_{\rm s}}{(r/r_{\rm s})(r/r_{\rm s} + 1)^2}, \nonumber \\
\Sigma(x) = 2\rho_{\rm s} x_{\rm s}\xi_0 \frac{1 - \mathcal{G}(x/x_{\rm s})}{(x/x_{\rm s})^2 - 1},
}
where $\rho_{\rm s}$ and $r_{\rm s}$ respectively denote a characteristic density at the center of the halo and a typical scale that characterizes the halo, and
\bae{
\mathcal{G}(a) = \begin{cases}
\frac{1}{\sqrt{a^2-1}}\text{arctan}(\sqrt{a^2-1}) & \text{for} ~a \geq 1, \nonumber \\
\frac{1}{\sqrt{1 - a^2}}\text{arctanh}(\sqrt{1 - a^2}) & \text{for} ~a \leq 1.
\end{cases}
}
We define $x_{\rm s} \equiv r_{\rm s}/\xi_0$. For simplicity, we assume $\rho_{\rm s}$ as
\bae{\rho_{\rm s} = \frac{\Sigma_{\rm cr}}{4x^3_{\rm s}\xi_0}\,,\label{eq:rhos_NFW}}
so that $\xi_0$ is equivalent to the Einstein radius of the system.
The deflection potential is given by
\bae{
\psi(x) = \frac{1}{2}\left[\log^2{\left(\frac{x/x_{\rm s}}{2}\right)} + \left((x/x_{\rm s})^2 -1\right)\mathcal{G}^2(x/x_{\rm s})\right].
} 
To obtain the amplification factor in the \ac{NFW} model, we need to solve Eq.~(\ref{eq: magnification_f}) numerically. In addition to $\xi_0$ and $y$, the amplification factor depends on $r_{\rm s}$.
The frequency dependence of the amplification factor for $y=0.3,~x_{\rm s} = 0.3$ is shown by the green dotted-dashed line in Fig.~\ref{fig: ampfactor_SIS_CIS_NFW}.

\section{Fisher forecasts of multi-band observation}
\label{sec: fisher}

In this section, we evaluate the statistical error in the determination of the parameters characterizing the lens system, e.g., the density profile of the dark matter halo and the impact parameter, by carrying out the Fisher analysis in a multi-band detection of the single \ac{GW} source. As the multi-band observation, we consider the two different combinations of \ac{GW} detectors: \ac{DECIGO} and \ac{ET}, and B-\ac{DECIGO} and \ac{ET}, covering $0.1$-$100$ Hz. To take full advantage of the benefits of multi-band observation, we employ a lens system in which the \ac{WO} regime can be observed by \ac{DECIGO} and B-\ac{DECIGO} and the \ac{GO} regime by \ac{ET}.

\subsection{Detectors}
   
\ac{ET}~\cite{Punturo:2010zz} is the third-generation ground-based detector planned to be built in the $2030$s. \ac{ET} covers the frequency $1$-$100$ Hz with the strain sensitivity $10^{-24}$ at the depth of the sensitivity curve, overwhelming the current detectors \ac{LIGO}, VIRGO, and KAGRA.
It is expected that
\ac{ET} can detect $100$ compact binaries as \ac{GW} sources per year with typical ${\rm SNR} \sim {\cal O}(100)$~\cite{Branchesi:2023mws}.
We employ the \ac{ET}-D design~\cite{Hild:2010id} for our analysis. We assume that \ac{ET} is located at the position on the Earth where the 
VIRGO detector is based on. 
\ac{DECIGO} is a Japan-led spece-borne interferometer being sensitive to \acp{GW} from 0.01 Hz to 10Hz. It is advantageous to observe the early inspiral part of \acp{GW} from stellar mass binary black holes.
B-\ac{DECIGO}~\cite{Kawamura:2018esd} is a pathfinder mission that will demonstrate various key techniques of \ac{DECIGO}. It still has several interesting science targets, e.g., foreground study, estimating the spin of the binary black holes.

\subsection{lens system and lensed \ac{GW} signal}
 
As we have mentioned, to take full advantage of the benefits of multi-band observation, we employ the lens system in which the \ac{WO} regime of the lensed \acp{GW} can be observed by \ac{DECIGO} and B-\ac{DECIGO} and the \ac{GO} regime by \ac{ET}.
Specifically, as a fiducial source, we consider the black hole binary with the primary mass $m_1 = 30 M_{\odot}$ and the secondary mass $m_2 = 20 M_{\odot}$, locating at a distance $11.2~\rm Gpc$ $(z_{\rm S} = 1.5)$ away from us.
Such a source is  
useful for our purpose because \ac{DECIGO} and B-\ac{DECIGO} observe the inspiral signal, and \ac{ET} observes from the late inspiral to the post-merger phase of the same \ac{GW} source. We found that the unlensed event will be detected with $\rm{SNR} \simeq 13$ for B-\ac{DECIGO},  $\rm{SNR} \simeq 163$ for \ac{DECIGO}, and $\rm{SNR} \simeq 23$ for the \ac{ET} observations.
We use the IMRPhenomD waveform template for an unlensed signal which is applied for the non-precessing model~\cite{Husa:2015iqa, Khan:2015jqa}. 
As the lens object, we consider the \ac{SIS}, \ac{CIS}, and \ac{NFW} models. As fiducial values of the parameters, the redshifted lens mass is taken to be $M_{{\rm L}z} = 3 \times 10^3 M_\odot$, the distance to the lens object is $6.8$~Gpc ($z_{\rm L}=1.0$), and the impact parameter, which is the source position on the lens plane, is set to be $y=0.3$ for the three models.  Note that $\xi_0$ is determined by Eq.~\eqref{eq:xi0_to_M_Lz} given $M_{{\rm L}z}$, $z_{\rm S}$, and $z_{\rm L}$. Then Eq.~\eqref{eq:v2_SIS}, Eq.~\eqref{eq:rho0_CIS}, and Eq.~\eqref{eq:rhos_NFW}, determines the amplitude of the density of profile for \ac{SIS}, \ac{CIS}, and \ac{NFW}. respectively. We assume $x_{\rm c} $ for \ac{CIS} and $x_{\rm s}$ for \ac{NFW} as $x_c = 0.3$ and $x_s = 0.3$, respectively.
The unlensed and lensed signals are numerically calculated by the public packages $\rm PyCBC$~\cite{nitz2024pycbc} and $\rm GLoW$ ~\cite{Villarrubia-Rojo:2024xcj}.

\begin{figure}[h] 
	\centering
	\includegraphics[width=9cm]{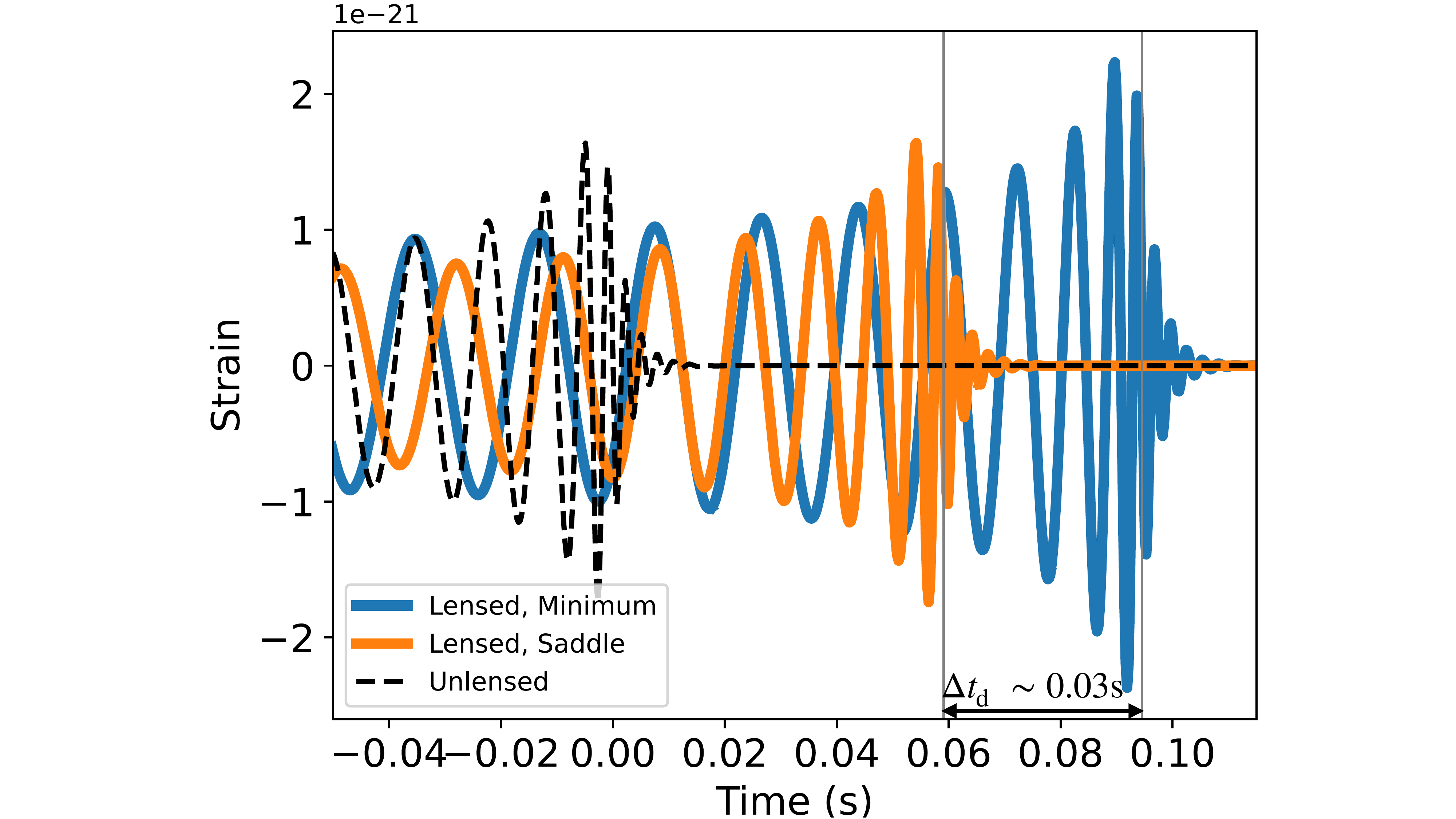}
	\caption{Unlensed and lensed waveforms in the time domain. We assume the \ac{SIS} lens with the fiducial values of the parameters described in Table~\ref{tab:parameters}. The black dashed line represents the unlensed signal. The blue and orange curves denote the lensed signal, which passes the minimum and the saddle points of the Fermat potential, respectively. We used the public code $\rm PyCBC$~\cite{nitz2024pycbc} for the plot.}
	\label{fig: ht_ET_DECIGO_SIS}
\end{figure}

\begin{figure}[h] 
	\centering
	\includegraphics[width=9cm]{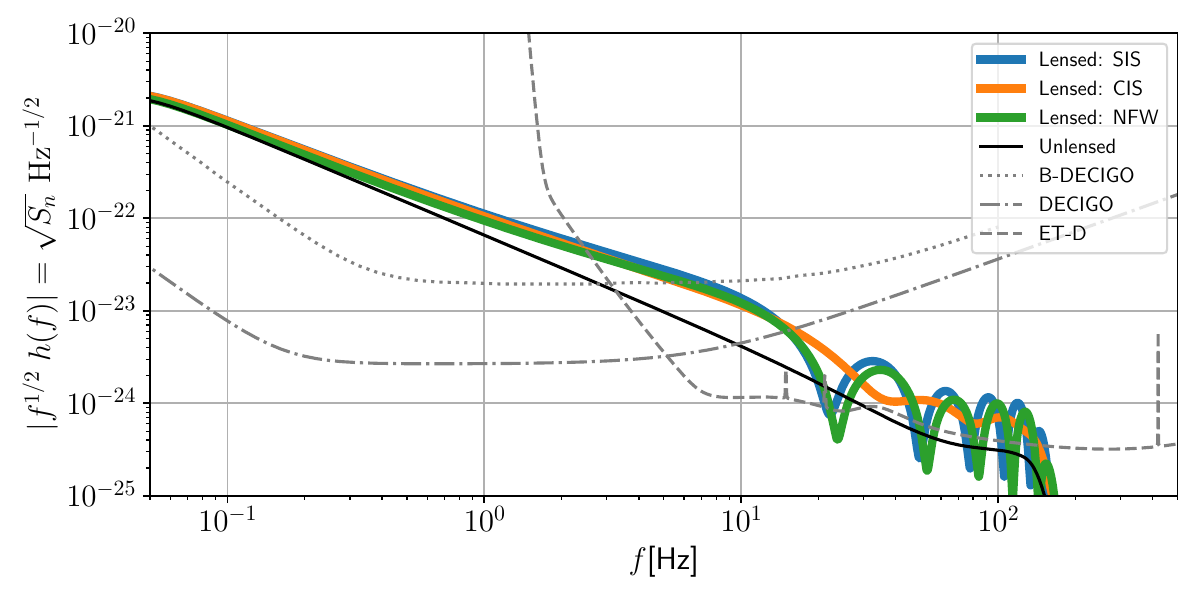}
	\caption{Unlensed (thin black) and lensed waveform in the Fourier domain for the \ac{SIS} (thick blue), \ac{CIS} (thick orange), and \ac{NFW} (thick green) models. We assume the fiducial values of the parameters described in Table~\ref{tab:parameters}. 
    The gray dotted, dash-dotted, and dashed curves respectively represent the B-\ac{DECIGO}~\cite{Nakamura:2016hna}, \ac{DECIGO}~\cite{Yagi:2011wg}, and \ac{ET}-D~\cite{Hild:2010id} sensitivity curves.}
	\label{fig: hf_ET_DECIGO_SIS_CIS_NFW}
\end{figure}

In Fig.~\ref{fig: ht_ET_DECIGO_SIS}, for the \ac{SIS} model with the fiducial parameters, we show the unlensed (thin black dashed curve) and the lensed (blue and orange solid curves) waveforms in the time domain. The blue and orange curves correspond to the signals that pass through the Fermat potential's minimum and saddle points, respectively.
In fact, in this setup, it is expected that the ring-down phase of the second lensed signal (blue) will arrive $\Delta t_{\rm d} \sim 0.03~\rm s$ after that of the first lensed signal (orange) is detected.
In Fig.~\ref{fig: hf_ET_DECIGO_SIS_CIS_NFW},
we show the unlensed (black solid curve) and the lensed \ac{GW} signals for three lens models: \ac{SIS} (thick blue), \ac{CIS} (thick orange) and \ac{NFW} (thick green).
We also show the planned sensitivity curves of \ac{ET}-D~\cite{Hild:2010id} (thin black Dashed), B-\ac{DECIGO}~\cite{Nakamura:2016hna} (thin black dotted), and \ac{DECIGO}~\cite{Yagi:2011wg} (thin black dash-dotted).
It is expected that in this setup,
the dark matter halo with the \ac{SIS} profile can be detected with $\rm{SNR} \simeq 21$ for B-\ac{DECIGO},  $\rm{SNR} \simeq 224$ for \ac{DECIGO}, and $\rm{SNR} \simeq 41$ for the \ac{ET} observations.
Similarly, the \ac{CIS} (\ac{NFW}) halo model can be detected with $\rm{SNR} \simeq 20~(21)$ for B-\ac{DECIGO},  $\rm{SNR} \simeq 217~(196)$ for \ac{DECIGO}, and $\rm{SNR} \simeq 36~(36)$ for the \ac{ET} observations.

\begin{table*}[t]
\vspace{0.5cm}
\centering
\caption{List of source and lens parameters used in this work. Fiducial values are adopted in our analysis.}
\vspace{0.5cm}
\resizebox{0.6\textwidth}{!}{%
\begin{tabular}{lll}
\hline\hline
Parameter & Description & Fiducial value \\
\hline
$\ln{\mathcal{M}_{\rm c}}$ & Redshifted chirp mass & $3.972$ \\
$\ln{\eta}$ & Symmetric mass ratio & $-1.427$ \\
$\ln{M_{\rm Lz}}$ &  Redshifted lens mass & $8.006$ \\
$y$ & Impact parameter & $0.3$ \\
$x_{\rm c}$ & Dimensionless core radius (CIS) & $0.3$ \\
$x_{\rm s}$ & Dimsensionless core radius (NFW) & $0.3$ \\
$d_{\rm L}$ [Gpc] & Luminosity distance & $11.2$ \\
$t_{\rm c}$ [sec] & Coalescence time & $0.004$ \\
$\phi_{\rm c}$ [rad] & Coalescence phase & $0.3$ \\
RA [rad] & Right ascension angle & $\pi/6$ \\
DEC [rad] & Declination angle & $\pi/4$ \\
$\theta_{\rm L}$ [rad] & Polar angle of orbital angular momentum & $\pi/12$ \\
$\phi_{\rm L}$ [rad] & Azimuthal angle of orbital angular momentum & $\pi/6$ \\
\hline
\end{tabular}%
}
\label{tab:parameters}
\vspace{1.0cm}
\end{table*}
%%%
%%%

\subsection{Fisher analysis}
We summarize the formalism of the Fisher analysis.
The observed time series data $s$ can be expressed as the sum of the true \ac{GW} signal $h_0$ and stationary Gaussian noise $n$ as
\bae{
s = h_0 + n.
}
The likelihood \( p(s|\bm{\theta}) \) quantifies the probability of observing the data \( s \), assuming the waveform model \( h(\bm{\theta}) \) with parameters \( \bm{\theta} \). 
The description of parameters $\bm{\theta}$ and the fiducial values in our setup are summarized in TABLE~\ref{tab:parameters}.

%%%

The likelihood is supposed to take the Gaussian form as ~\cite{Vallisneri:2007ev,Rodriguez:2013mla}
\bae{
\label{eq: likelihood}
p(s|\bm \theta) &\propto \exp\left[-\frac{(n, n)}{2}\right] \nonumber\\
& = \exp\left[-\frac{\left(s - h(\bm \theta), s - h(\bm \theta)\right)}{2}\right],
}
where $(a, b)$ is the inner product given by
\bae{
(a, b) \equiv 4 \Re \int^{f_\mathrm{max}}_{f_\mathrm{min}} \dd{f} \frac{\tilde{a}(f) \tilde{b}^*(f)}{S_n(f)}.
}
Here, $\tilde{a}$ and $\tilde{b}$ are the Fourier transformations in the frequency domain of the scalar functions $a$ and $b$, respectively. Around here, ${}^*$ denotes the complex conjugate. $S_n$ is the one-sided power spectral density of the noise, which is given by the cross-correlation as
\bae{
\label{eq: noise_power}
 \langle \tilde{n}^*(f^{\prime}) \tilde{n}(f) \rangle = \frac{1}{2}S_n(\abs{f})\delta(f-f^{\prime}),
}
where $\tilde{n}$ is the Fourier transformation of the intrinsic detector noise $n$. $\langle \cdots \rangle$ means averaging over the probability distribution of the noise. 
The Fisher information matrix is defined by the second derivative (the negative Hessian) of the likelihood as
\bae{
\Gamma_{ij} &\equiv - \langle \partial_i \partial_j \log p(s|\bm \theta) \rangle |_{\bm \theta = \bm \theta_0},\\ \nonumber
&=  \langle \partial_i \log p(s|\bm \theta)\partial_j \log p(s|\bm \theta) \rangle |_{\bm \theta = \bm \theta_0},
}
where $\partial_i$ means taking the derivative respect to the parameter $\theta_i$ and $h_0 = h(\bm{\theta}_0)$ denotes the true \ac{GW} signal. The above expression is equivalent to
\bae{
\Gamma_{ij} &= (\partial_i h, \partial_j h) \nonumber \\
&= 4 \Re \int^{f_\mathrm{max}}_{f_\mathrm{min}} \dd{f} \frac{\partial_i \tilde{h}(f, \bm{\theta}) \partial_j \tilde{h}^*(f,\bm{\theta})}{S_n(f)}.
}
We need to set the minimum and the maximum frequencies $[f_{\rm min},~f_{\max}]$ to carry out the integration in the Fisher matrix. We choose $[f_{\rm min}/\rm{Hz}, ~\mathit{f}_{\max}/\rm{Hz}] = [0.05,~20],$ and $[10,~200]$ for (B-)\ac{DECIGO} and \ac{ET} respectively.
For joint observation, we assume that the noises in different detectors are statistically independent. Then, we get the total Fisher matrix by
\begin{equation}
    \Gamma_{ij} = \qty[\Gamma^\mathrm{DECIGO} + \Gamma^\mathrm{ET}]_{ij}\,.
\end{equation}
We will evaluate the Fisher matrix by fixing parameters $\bm \theta = \{d_{\rm L}, t_{\rm c}, \phi_{\rm c}, \rm{RA}, \rm{DEC}, \bar{\theta}_{\rm L}, \bar{\phi}_{\rm L}\}$. The corner plots presented in the main text are generated by fixing these parameters for clarity. Comprehensive results including all model parameters are provided in Appendix~\ref{appendix: supplement}. We have verified that the impact of fixing these parameters on the accuracy of the inferred lens model parameters is limited.

%%%% ET x B-DECIGO
\begin{figure*}[t]
    \centering

    % First row: one figure
    \begin{subfigure}[t]{0.49\textwidth}
        \centering
        \includegraphics[width=1.125\textwidth]{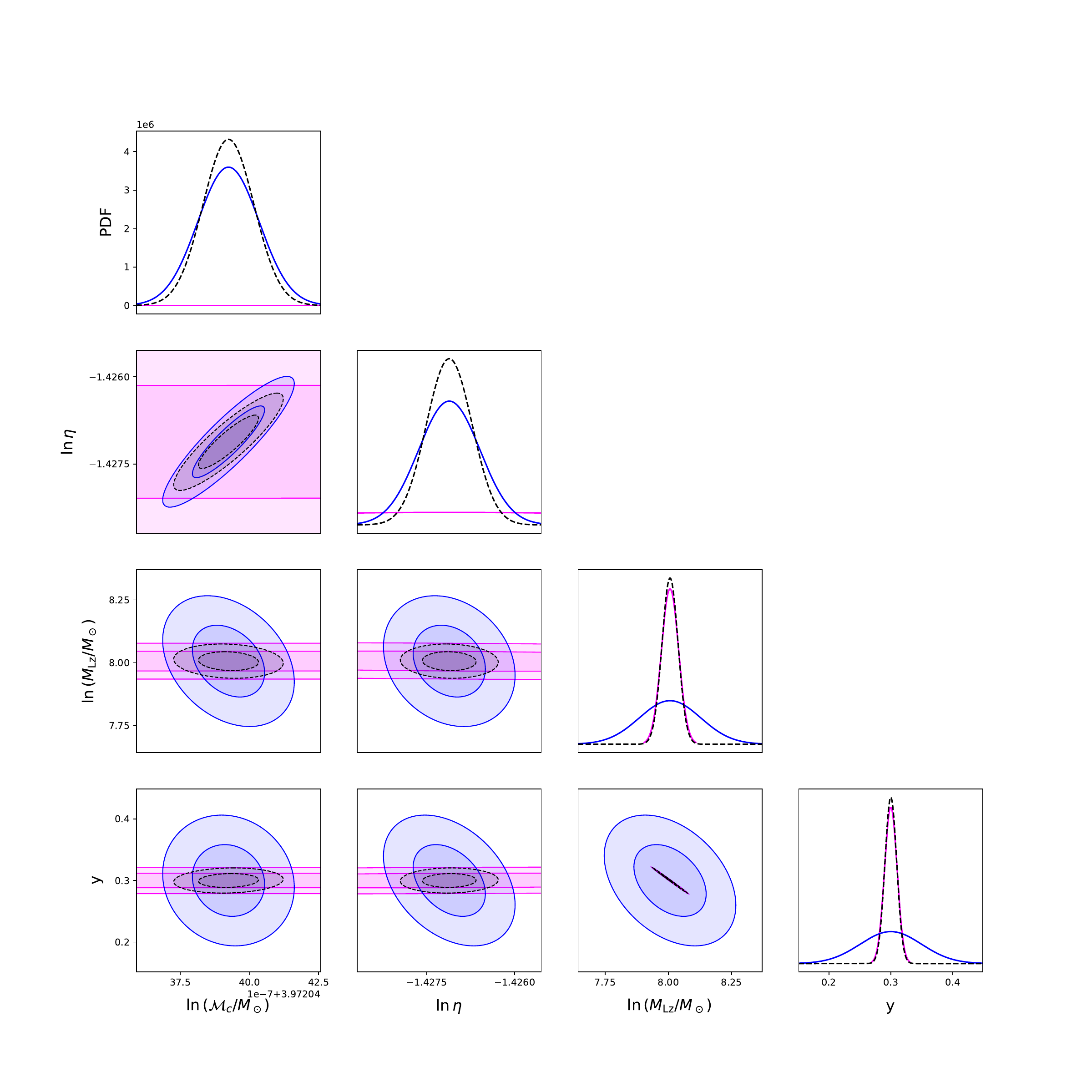}
        \caption{SIS lens model}
        \label{fig: Fisher_SIS_ET_BDECIGO}
    \end{subfigure}

    \vspace{1.5em}

    \begin{subfigure}[t]{0.49\textwidth}
        \centering
        \includegraphics[width=1.125\textwidth]{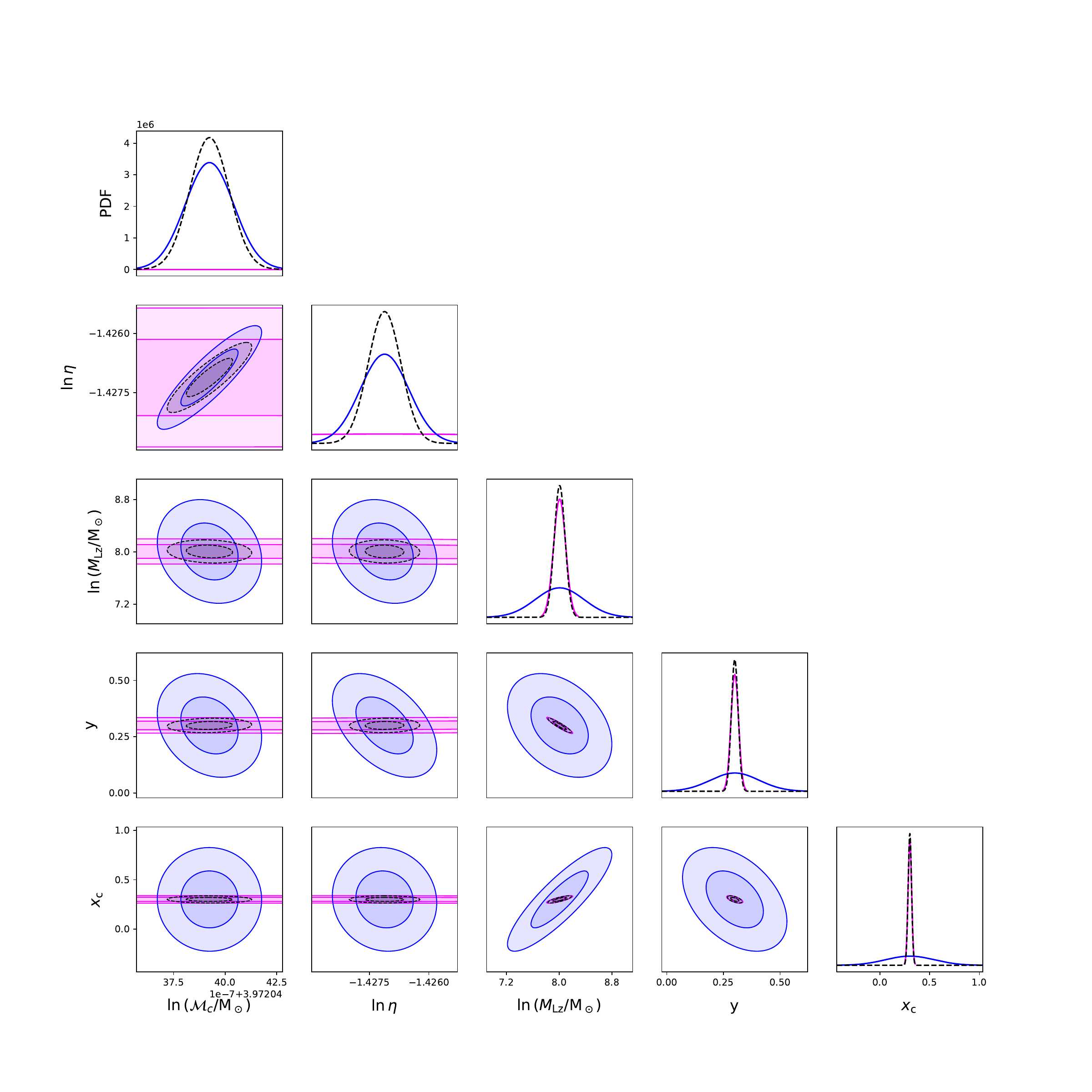}
        \caption{CIS lens model}
        \label{fig: Fisher_CIS_ET_BDECIGO}
    \end{subfigure}
    \hfill
    \begin{subfigure}[t]{0.49\textwidth}
        \centering
        \includegraphics[width=1.125\textwidth]{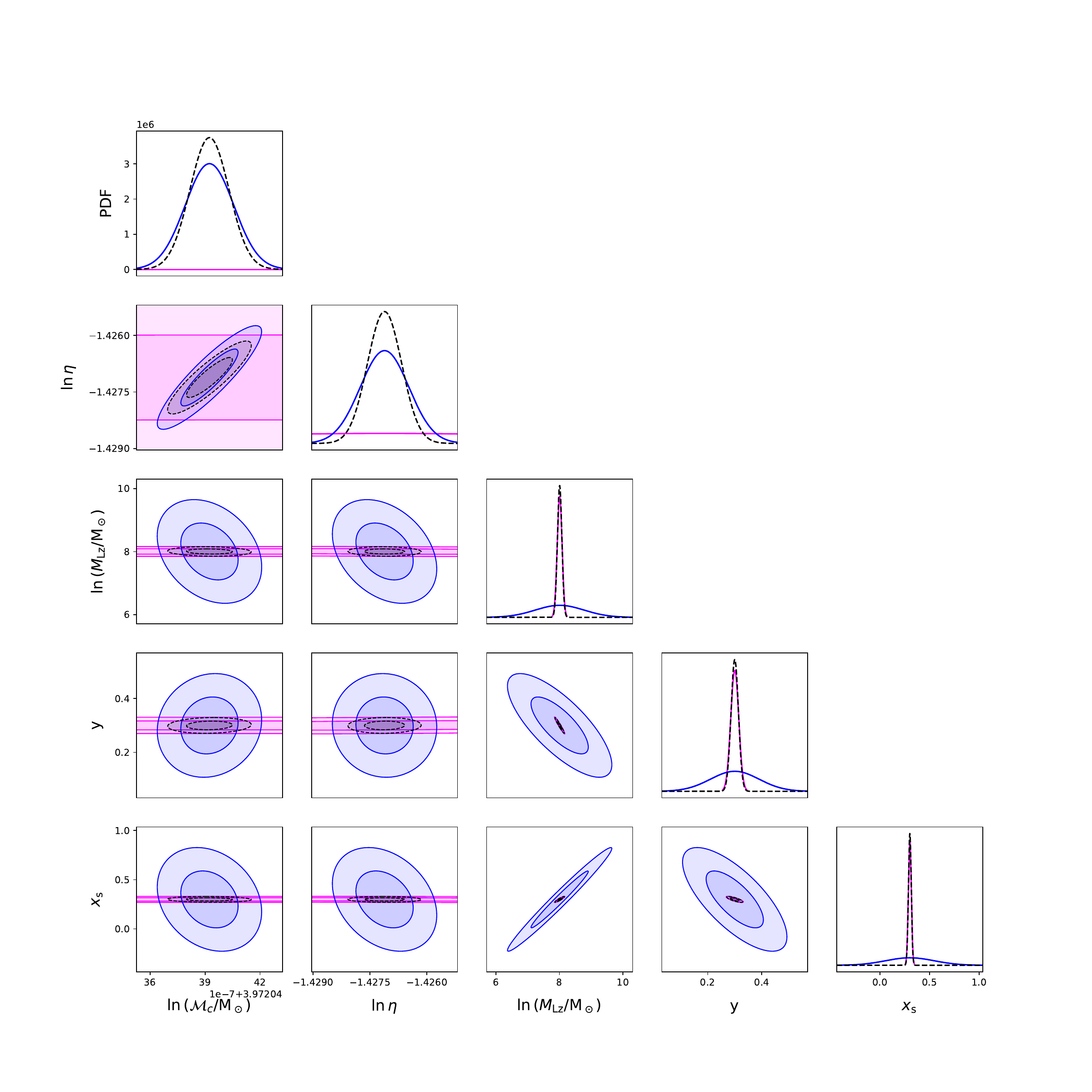}
        \caption{NFW lens model}
        \label{fig: Fisher_NFW_ET_BDECIGO}
    \end{subfigure}

    \caption{$50\%$ and $90\%$ error contours for B-\ac{DECIGO} and \ac{ET} in different lens models. Magenta and blue contours represent forecasts from \ac{ET} and B-\ac{DECIGO}, respectively. The black dashed contours show the joint analysis.}
    \label{fig: Fisher_all_lens_models_BDECIGO}
\end{figure*}

%%%% ET x DECIGO
\begin{figure*}[h]
    \centering

    \begin{subfigure}[t]{0.49\textwidth}
        \centering
        \includegraphics[width=1.125\textwidth, height=1.125\textwidth]{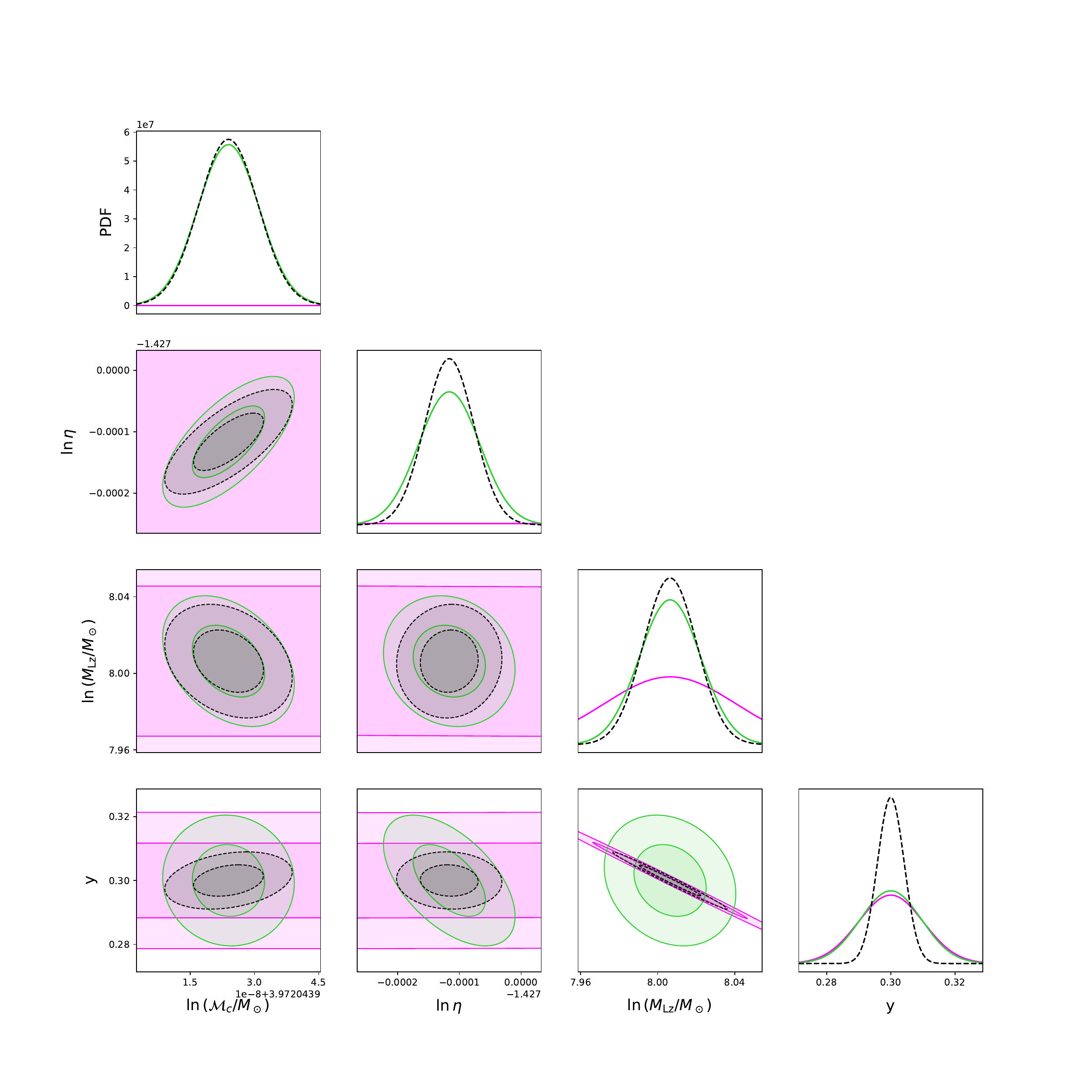}
        \caption{SIS lens model}
        \label{fig: Fisher_SIS_ET_DECIGO}
    \end{subfigure}

    \vspace{0.5em}

    \begin{subfigure}[t]{0.49\textwidth}
        \centering
        \includegraphics[width=1.125\textwidth, height=1.125\textwidth]{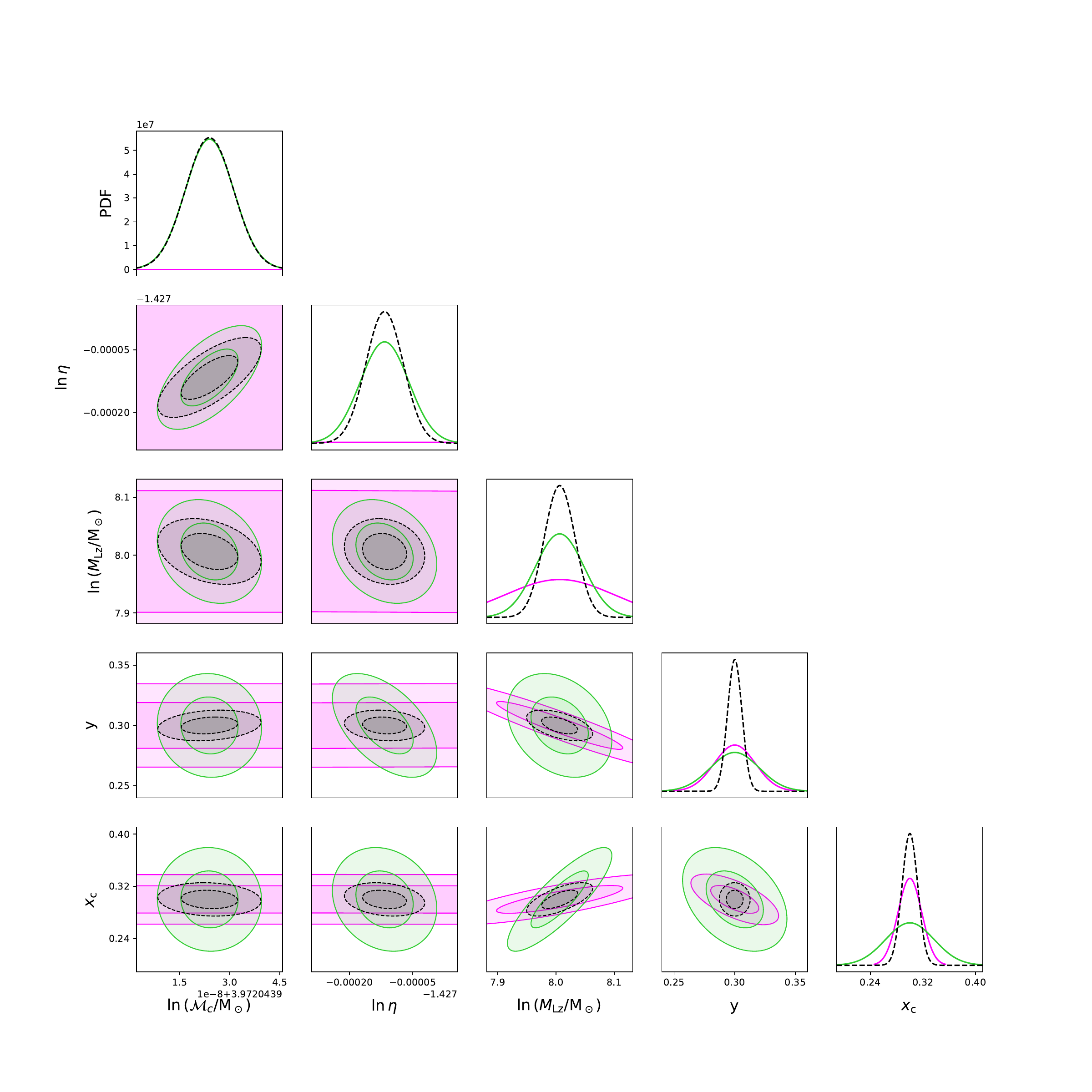}
        \caption{CIS lens model}
        \label{fig: Fisher_CIS_ET_DECIGO}
    \end{subfigure}
    \hfill
    \begin{subfigure}[t]{0.49\textwidth}
        \centering
        \includegraphics[width=1.125\textwidth, height=1.125\textwidth]{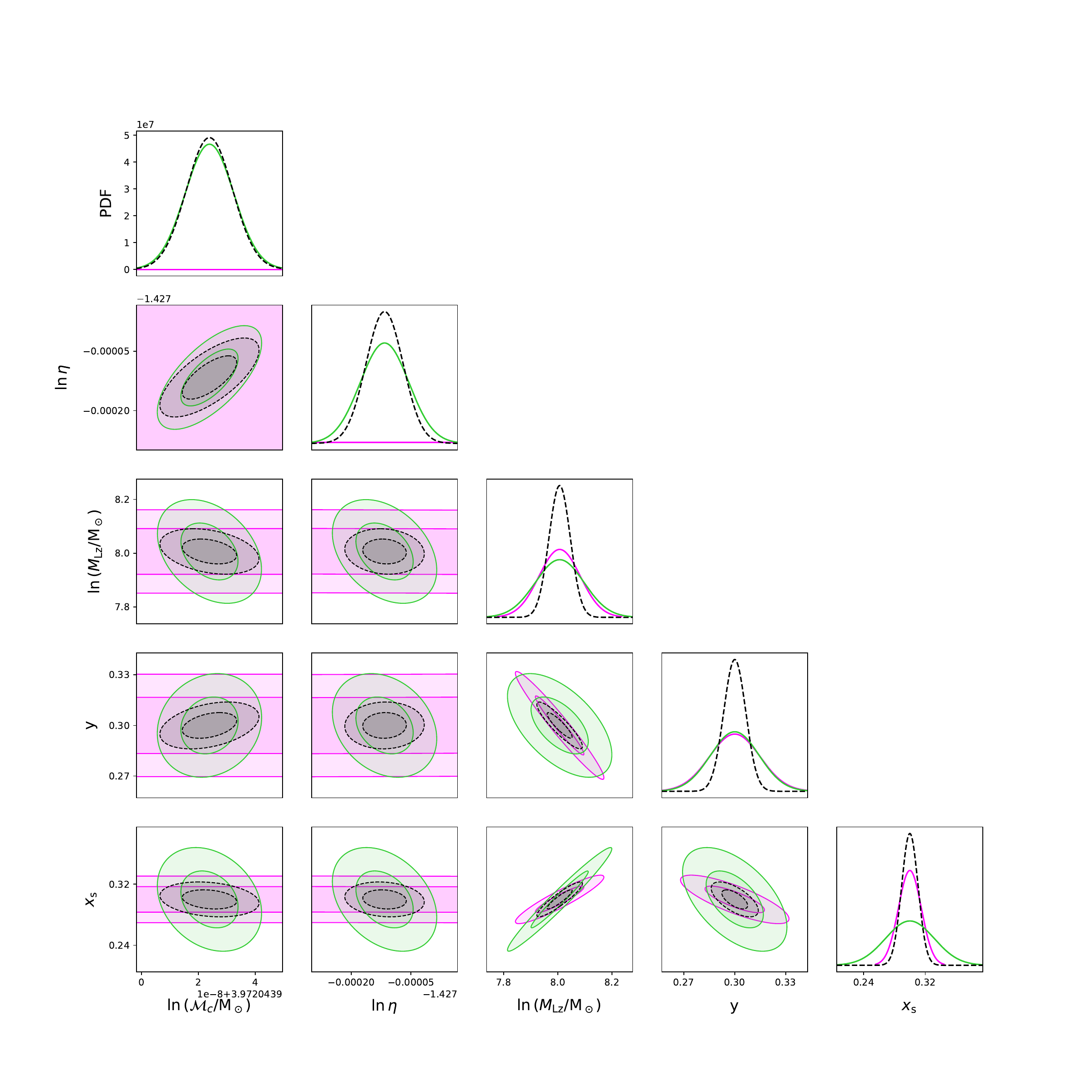}
        \caption{NFW lens model}
        \label{fig: Fisher_NFW_ET_DECIGO}
    \end{subfigure}

    \caption{ $50\%$ and $90\%$ error contours for \ac{DECIGO} and \ac{ET} in different lens models. Magenta and green contours represent forecasts from \ac{ET} and \ac{DECIGO}, respectively. The black dashed contours show the joint analysis.}
    \label{fig: Fisher_all_lens_models_DECIGO}
\end{figure*}

\begin{table*}[h]
\centering
\caption{The statistical error estimations for B-\ac{DECIGO} and \ac{ET} observations assuming different lens models.}
\label{table: error_all_ET_BDECIGO}

\makebox[\textwidth]{%
\begin{subtable}{0.95\textwidth}
\centering
\caption{SIS lens model}
\begin{tabular}{|c|c|c|c|c|}
\hline
$\rm{Detector}$ & $\Delta \ln{\mathcal{M}_{\rm{c}}}/\ln{\mathcal{M}_{\rm{c}}}~(\%)$ & $\Delta \ln{\eta}/\ln{\eta}~(\%)$ & $\Delta \ln{M_{\rm{Lz}}}/\ln{M_{\rm{Lz}}}~(\%)$ & $\Delta y/y~(\%)$ \\[1ex]
\hline
$\rm{ET}$ & 0.038 & 0.362 & 0.425 & 3.381  \\[1ex]
\hline
$\rm{BDECIGO}$ & $3\times10^{-6}$ & 0.037 & 1.514 & 16.50  \\[1ex]
\hline
$\rm{ET + BDECIGO}$ & $2\times10^{-6}$ & 0.027 & 0.397 & 3.177 \\[1ex]
\hline
\end{tabular}
\end{subtable}
}

\vspace{1em}

\makebox[\textwidth]{%
\begin{subtable}{0.95\textwidth}
\centering
\caption{CIS lens model}
\begin{tabular}{|c|c|c|c|c|c|}
\hline
$\rm{Detector}$ & $\Delta \ln{\mathcal{M}_{\rm{c}}}/\ln{\mathcal{M}_{\rm{c}}}~(\%)$ & $\Delta \ln{\eta}/\ln{\eta}~(\%)$ & $\Delta \ln{M_{\rm{Lz}}}/\ln{M_{\rm{Lz}}}~(\%)$ & $\Delta y/y~(\%)$ & $\Delta x_{\rm c} / x_{\rm c} ~(\%)$ \\[1ex]
\hline
$\rm{ET}$ & 0.043 & 0.408 & 1.152 & 5.621 & 6.022  \\[1ex]
\hline
$\rm{BDECIGO}$ & $3\times10^{-6}$ & 0.043 & 4.610 & 35.77 & 81.66  \\[1ex]
\hline
$\rm{ET + BDECIGO}$ & $2\times10^{-6}$ & 0.029 & 1.033 & 4.980 & 5.674 \\[1ex]
\hline
\end{tabular}
\end{subtable}
}

\vspace{1em}

\makebox[\textwidth]{%
\begin{subtable}{0.95\textwidth}
\centering
\caption{NFW lens model}
\begin{tabular}{|c|c|c|c|c|c|}
\hline
$\rm{Detector}$ & $\Delta \ln{\mathcal{M}_{\rm{c}}}/\ln{\mathcal{M}_{\rm{c}}}~(\%)$ & $\Delta \ln{\eta}/\ln{\eta}~(\%)$ & $\Delta \ln{M_{\rm{Lz}}}/\ln{M_{\rm{Lz}}}~(\%)$ & $\Delta y/y~(\%)$ & $\Delta x_{\rm s} / x_{\rm s} ~(\%)$ \\[1ex]
\hline
$\rm{ET}$ & 0.043 & 0.418 & 0.952 & 4.964 & 4.928  \\[1ex]
\hline
$\rm{BDECIGO}$ & $3\times10^{-6}$ & 0.045 & 9.571 & 29.95 & 81.76  \\[1ex]
\hline
$\rm{ET + BDECIGO}$ & $3\times10^{-6}$ & 0.032 & 0.875 & 4.555 & 4.661 \\[1ex]
\hline
\end{tabular}
\end{subtable}
}

\end{table*}

%%%%%%%%

%%%% ET x DECIGO %%%%
\begin{table*}[t]
\centering
\caption{The statistical error estimations for \ac{DECIGO} and \ac{ET} observations assuming different lens models.}
\label{table: error_all_lens_models}

% SIS lens model
\makebox[\textwidth]{%
\begin{subtable}{0.95\textwidth}
\centering
\caption{SIS lens model}
\label{table: error_SIS_ET_DECIGO}
\begin{tabular}{|c|c|c|c|c|}
\hline
$\rm{Detector}$ & $\Delta \ln{\mathcal{M}_{\rm{c}}}/\ln{\mathcal{M}_{\rm{c}}}~(\%)$ 
& $\Delta \ln{\eta}/\ln{\eta}~(\%)$ 
& $\Delta \ln{M_{\rm{Lz}}}/\ln{M_{\rm{Lz}}}~(\%)$ 
& $\Delta y/y~(\%)$ \\[1ex]
\hline
$\rm{ET}$ & 0.038 & 0.362 & 0.425 & 3.381 \\[1ex]
\hline
$\rm{DECIGO}$ & $1.804\times10^{-7}$ & 0.003 & 0.199 & 3.183 \\[1ex]
\hline
$\rm{ET + DECIGO}$ & $1.746\times10^{-7}$ & 0.003 & 0.173 & 1.393 \\[1ex]
\hline
\end{tabular}
\end{subtable}
}

\vspace{1em}

% CIS lens model
\makebox[\textwidth]{%
\begin{subtable}{0.95\textwidth}
\centering
\caption{CIS lens model}
\label{table: error_CIS_ET_DECIGO}
\begin{tabular}{|c|c|c|c|c|c|}
\hline
$\rm{Detector}$ & $\Delta \ln{\mathcal{M}_{\rm{c}}}/\ln{\mathcal{M}_{\rm{c}}}~(\%)$ 
& $\Delta \ln{\eta}/\ln{\eta}~(\%)$ 
& $\Delta \ln{M_{\rm{Lz}}}/\ln{M_{\rm{Lz}}}~(\%)$ 
& $\Delta y/y~(\%)$ 
& $\Delta x_{\rm c} / x_{\rm c} ~(\%)$ \\[1ex]
\hline
$\rm{ET}$ & 0.043 & 0.408 & 1.152 & 5.621 & 6.022 \\[1ex]
\hline
$\rm{DECIGO}$ & $1.837\times10^{-7}$ & 0.004 & 0.522 & 6.685 & 12.329 \\[1ex]
\hline
$\rm{ET + DECIGO}$ & $1.816\times10^{-7}$ & 0.003 & 0.331 & 1.974 & 3.973 \\[1ex]
\hline
\end{tabular}
\end{subtable}
}

\vspace{1em}

% NFW lens model
\makebox[\textwidth]{%
\begin{subtable}{0.95\textwidth}
\centering
\caption{NFW lens model}
\label{table: error_NFW_ET_DECIGO}
\begin{tabular}{|c|c|c|c|c|c|}
\hline
$\rm{Detector}$ & $\Delta \ln{\mathcal{M}_{\rm{c}}}/\ln{\mathcal{M}_{\rm{c}}}~(\%)$ 
& $\Delta \ln{\eta}/\ln{\eta}~(\%)$ 
& $\Delta \ln{M_{\rm{Lz}}}/\ln{M_{\rm{Lz}}}~(\%)$ 
& $\Delta y/y~(\%)$ 
& $\Delta x_{\rm s} / x_{\rm s} ~(\%)$ \\[1ex]
\hline
$\rm{ET}$ & 0.043 & 0.418 & 0.952 & 4.964 & 4.928 \\[1ex]
\hline
$\rm{DECIGO}$ & $2.154\times10^{-7}$ & 0.004 & 1.121 & 4.771 & 10.539 \\[1ex]
\hline
$\rm{ET + DECIGO}$ & $2.046\times10^{-7}$ & 0.003 & 0.491 & 2.156 & 3.546 \\[1ex]
\hline
\end{tabular}
\end{subtable}
}

\end{table*}
\clearpage

\subsection{B-DECIGO and ET}
\label{ssec:BD_ET_results}
Let us show the results assuming the B-\ac{DECIGO} and the \ac{ET} observations in Fig.~\ref{fig: Fisher_all_lens_models_BDECIGO}. The blue contours correspond to the forecasts of B-\ac{DECIGO} and the magenta contours show the estimates of \ac{ET}. The black dashed contours are the results from the joint analysis of B-\ac{DECIGO} and \ac{ET}.
Figure~\ref{fig: Fisher_SIS_ET_BDECIGO},  Fig.~\ref{fig: Fisher_CIS_ET_BDECIGO}, and Fig.~\ref{fig: Fisher_NFW_ET_BDECIGO} show 50~\% and 90~\% error contours obtained by assuming the \ac{SIS}, the \ac{CIS}, and the \ac{NFW} lens models, respectively. 
As can be seen in these figures, B-\ac{DECIGO} gives much tighter constraints on the chirp mass and the symmetric mass ratio than \ac{ET}. On the other hand, the parameters of the lens, $M_{{\rm L}z}$ and $y$, would be determined by \ac{ET} only. This tendency is common for all three lens models. Although there are slight improvements in the statistical errors of lens parameters by use of the multi-band observation with B-\ac{DECIGO} and \ac{ET}, we find that this combination of B-\ac{DECIGO} and \ac{ET} is not so powerful in weakening the degeneracy between parameters. We summarize the estimated statistical errors of parameters for B-\ac{DECIGO} and \ac{ET} in Table~\ref{table: error_all_ET_BDECIGO}.

\subsection{DECIGO and ET band}\label{ssec:D_ET_results}

In Figure~\ref{fig: Fisher_all_lens_models_DECIGO}, we show the Fisher forecasts assuming the \ac{DECIGO} and the \ac{ET} observations. The green contours correspond to the forecasts of \ac{DECIGO} and the magenta contours show the estimates of \ac{ET}. The black dashed contours are the results from the joint analysis of \ac{DECIGO} and \ac{ET}.
Fig.~\ref{fig: Fisher_SIS_ET_DECIGO},  Fig.~\ref{fig: Fisher_CIS_ET_DECIGO}, and Fig.~\ref{fig: Fisher_NFW_ET_DECIGO} show 50~\% and 90~\% error contours obtained by assuming the \ac{SIS}, the \ac{CIS}, and the \ac{NFW} lens models,  respectively.
Although \ac{DECIGO} could outperform \ac{ET} in the accuracy of the parameter estimation for almost all of the parameters, including the lens parameters, 
the significant reduction of the errors takes place for the lens parameters when combining the two detectors. In fact, as shown in Table~\ref{table: error_SIS_ET_DECIGO}, the combination of \ac{ET} and \ac{DECIGO} can improve the determination accuracy of the impact parameter $y$ from $3.381\%~(3.183\%)$ for \ac{ET} (\ac{DECIGO}) to  $1.393\%$.
This indicates quantitatively that the multi-band observation reduces errors in the impact parameter by about $59\%$ and $56\%$, compared to observing only by ET and DECIGO, respectively. 
As for the \ac{CIS} lens model,
it is noticed that the degeneracy between $M_{\mathrm{Lz}}$ and $y$, and also between  $M_{\mathrm{Lz}}$ and $x_{\mathrm{c}}$ are mitigated by combining the \ac{DECIGO} and the \ac{ET} observations (see the correlation coefficient maps in Appendix.~\ref{appendix: supplement}).
The estimated error for the CIS model is in Table.~\ref{table: error_CIS_ET_DECIGO}.
As can be seen in this table,
the joint observation of \ac{ET} and \ac{DECIGO} reduces  the statistical error in the lens mass $M_{{\rm L}z}$
by about $71\%$ and $58\%$ compared to single observations in \ac{ET} and \ac{DECIGO}, respectively.
Similarly, the error in the impact parameter $y$ would be reduced by about $65\%$ and $70\%$.
The parameter $x_{\rm c}$, which characterizes the size of the core in the \ac{CIS} model,
is also improved by $34\%$ and $68\%$
compared to the individual observation in \ac{ET} and DECIGO.
Additionally, we can realize the improvement of the statistical error for the model parameters of the \ac{NFW} model, as shown in Table~\ref{table: error_NFW_ET_DECIGO}.

\section{Discussion and Conclusion}

The more attention \ac{GW} astronomy attracts, the more actively studies on the \ac{GL} effect of the \acp{GW} have been discussed. In particular, there are many studies on lensed \acp{GW} as a tool to inspect small-scale structures composed of DM in the Universe, confronting the CDM paradigm in cosmology. The \ac{WO} of \acp{GW} has been considered as a new way of probing small-mass DM halos, whereas the parameter degeneracy among the lens parameters that describe the DM halo has been found.
To clarify whether the joint analysis of the \ac{GO} and \ac{WO} effects of GWs sourced by a small-mass DM halo can be advantageous to constrain the DM properties as well as source parameters,
we investigated the \ac{GW} lensing in the multi-band observation assuming \ac{ET} and (B-)\ac{DECIGO}, and we discussed how much the determination accuracy of the lens model parameters is improved by the joint analysis using the Fisher information matrix.
For simplicity, we assumed the non-precessing/non-spinning binary wave source with the primary mass $m_1 = 30 M_{\odot}$, the secondary mass $m_2 = 20 M_{\odot}$ locating at the distance $11.2~\rm Gpc$ $(z = 1.5)$ away from us.
In fact, such an unlensed event is expected to be detectable with $\rm{SNR} \simeq 13, 163,$ and $23$ for B-\ac{DECIGO}, \ac{DECIGO}, and \ac{ET} in the future. We gave Fisher forecasts assuming this event is lensed by the DM halo with the mass of $3\times 10^{3}M_\odot$ at  $6.8~\rm{Gpc}$ $(z = 1.0)$. As a fiducial value of the impact parameter $y$, which characterizes the relative position of the source to the lens object on the source plane, we assumed $y=0.3$. We employ three halo density profile models: \ac{SIS}, \ac{CIS}, and \ac{NFW}. 
The unlensed and lensed waveforms in the Fourier domain are displayed in Fig.~\ref{fig: hf_ET_DECIGO_SIS_CIS_NFW}. 

In this setup, we first investigated the precision of determination for the combination of the \ac{ET} and the B-\ac{DECIGO} observations. 
While 
B-\ac{DECIGO} gives much tighter constraints on the chirp mass and the symmetric mass ratio than \ac{ET}, the statistical errors for the lens model parameters are found to be determined by \ac{ET}. This tendency is common for all three lens models. Although there are slight improvements in the statistical errors of lens parameters by the combination of B-\ac{DECIGO} and \ac{ET}, for this combination, the observation of the \ac{GO} and \ac{WO} effects by multi-band observations cannot be advantageous in the parameter estimation.

We also discussed the accuracy of determination for the \ac{ET} and the \ac{DECIGO} observations.
For the \ac{SIS} model, although almost all of the parameters can be determined solely by \ac{DECIGO}, the improvement of the error can be seen in the parameter $y$ by the combination of \ac{ET} and \ac{DECIGO}. 
Quantitatively, as shown in Table~\ref{table: error_SIS_ET_DECIGO}, combining \ac{ET} and \ac{DECIGO} reduces the errors in the impact parameter by about $59\%$ and $56\%$, compared to observing only by \ac{ET} and \ac{DECIGO}, respectively. 
We interpret how the degeneracy is mitigated as follows. In the high-frequency regime, the magnification is determined by the geometrical configuration derived from the lens equation, involving the impact parameter $y$, $x_{c}$, and $x_s$.
In contrast, in the low-frequency regime where the \ac{WO} effects dominate, the magnification is less sensitive to the impact parameter due to the interference, leaving $M_{\mathrm{Lz}}$ determines the amplitude.
Therefore, multi-band observations can complement the information from the lens and thus improve the determination of lens parameters. 
 
Remarkably, for the \ac{CIS} and \ac{NFW} models, which have the additional lens parameter $x_{\rm c}$ and $x_{\rm s}$, respectively, the combination of \ac{ET} and \ac{DECIGO} outperforms in parameter estimation by reducing the statistical errors compared
%%%%%%%%%%%%%%%%%%%%%%%%%%%%%
%
%%%%%%%%%%%%%%%%%%%%%%%%%%%%%
to either the \ac{GO}-only or \ac{WO}-only detection by the individual detectors, 
relaxing the degeneracy between the three parameters, $y$, $M_{{\rm L}z}$, and $x_{\rm c}$/$x_{\rm s}$. As shown in Tables~\ref{table: error_CIS_ET_DECIGO} and~\ref{table: error_NFW_ET_DECIGO}, the statistical error for $x_{\rm c}$ ($x_{\rm s}$) for the \ac{CIS} (\ac{NFW}) model can be improved by about $34\%$ (28\%) and $68\%$ (66\%), compared to the individual observations in \ac{ET} and \ac{DECIGO}, respectively.
Thus, our analysis has revealed that the multi-band observation by \ac{ET} and (B-)\ac{DECIGO} can improve the determination accuracy of the lens model parameters.
In particular, the use of \ac{DECIGO}
potentially allows us to take full advantage of the multi-band observation, measuring both of the \ac{GO} and \ac{WO} effects that appear in the lensed \acp{GW}, and to mitigate the degeneracy between the lens model parameters.

In fact, in this paper, we demonstrated the multi-band analysis based on simple assumptions, for example, non-processing/spinning binaries, the simultaneous observation by \ac{ET} and (B-)\ac{DECIGO} of the same lensed event, and the simple lens models such as \ac{SIS}, \ac{CIS}, and \ac{NFW} models.
We would like to point out that a more detailed analysis is needed to reach a more robust detectability. For example, we should consider the population and the event rate for the source and the lens object that can be simultaneously observed in \ac{ET} and (B-) \ac{DECIGO} detectors.
In addition, it should be important to
discuss the lensing signal on general lens models such as the Pseudo-Jaffe ellipsoid model, elliptical \ac{SIS}, \ac{GSIS}, and so on. 
We also take into account that the large correlation between spin and mass increases statistical uncertainties~\cite{Cutler:1994ys, Lee:2022jpn}. We leave these points for future work.

\label{sec: conclusion}

\section{Acknowledgements}
We thank Takahiro Nishimichi and Soichiro Morisaki for fruitful discussions.
RI is supported by the Japan Society for the Promotion of Science
(JSPS) Grants-in-Aid for Scientific Research (KAKENHI) Grant No. JP25KJ1442.
SA is supported by the Japan Society for the Promotion of Science
(JSPS) Grants-in-Aid for Scientific Research (KAKENHI) Grant No. JP24K17045.
KTA is currently affiliated with Fujitsu Limited, Kanagawa, Japan.
TSY is supported by JSPS KAKENHI Grant Numbers JP23K13099 and JP23H04502.

\bibliography{main}

\begin{thebibliography}{64}%
\makeatletter
\providecommand \@ifxundefined [1]{%
 \@ifx{#1\undefined}
}%
\providecommand \@ifnum [1]{%
 \ifnum #1\expandafter \@firstoftwo
 \else \expandafter \@secondoftwo
 \fi
}%
\providecommand \@ifx [1]{%
 \ifx #1\expandafter \@firstoftwo
 \else \expandafter \@secondoftwo
 \fi
}%
\providecommand \natexlab [1]{#1}%
\providecommand \enquote  [1]{``#1''}%
\providecommand \bibnamefont  [1]{#1}%
\providecommand \bibfnamefont [1]{#1}%
\providecommand \citenamefont [1]{#1}%
\providecommand \href@noop [0]{\@secondoftwo}%
\providecommand \href [0]{\begingroup \@sanitize@url \@href}%
\providecommand \@href[1]{\@@startlink{#1}\@@href}%
\providecommand \@@href[1]{\endgroup#1\@@endlink}%
\providecommand \@sanitize@url [0]{\catcode `\\12\catcode `\$12\catcode `\&12\catcode `\#12\catcode `\^12\catcode `\_12\catcode `\%12\relax}%
\providecommand \@@startlink[1]{}%
\providecommand \@@endlink[0]{}%
\providecommand \url  [0]{\begingroup\@sanitize@url \@url }%
\providecommand \@url [1]{\endgroup\@href {#1}{\urlprefix }}%
\providecommand \urlprefix  [0]{URL }%
\providecommand \Eprint [0]{\href }%
\providecommand \doibase [0]{https://doi.org/}%
\providecommand \selectlanguage [0]{\@gobble}%
\providecommand \bibinfo  [0]{\@secondoftwo}%
\providecommand \bibfield  [0]{\@secondoftwo}%
\providecommand \translation [1]{[#1]}%
\providecommand \BibitemOpen [0]{}%
\providecommand \bibitemStop [0]{}%
\providecommand \bibitemNoStop [0]{.\EOS\space}%
\providecommand \EOS [0]{\spacefactor3000\relax}%
\providecommand \BibitemShut  [1]{\csname bibitem#1\endcsname}%
\let\auto@bib@innerbib\@empty
%</preamble>
\bibitem [{\citenamefont {Riess}\ \emph {et~al.}(1998)\citenamefont {Riess} \emph {et~al.}}]{SupernovaSearchTeam:1998fmf}%
  \BibitemOpen
  \bibfield  {author} {\bibinfo {author} {\bibfnamefont {A.~G.}\ \bibnamefont {Riess}} \emph {et~al.} (\bibinfo {collaboration} {Supernova Search Team}),\ }\bibfield  {title} {\bibinfo {title} {{Observational evidence from supernovae for an accelerating universe and a cosmological constant}},\ }\href {https://doi.org/10.1086/300499} {\bibfield  {journal} {\bibinfo  {journal} {Astron. J.}\ }\textbf {\bibinfo {volume} {116}},\ \bibinfo {pages} {1009} (\bibinfo {year} {1998})},\ \Eprint {https://arxiv.org/abs/astro-ph/9805201} {arXiv:astro-ph/9805201} \BibitemShut {NoStop}%
\bibitem [{\citenamefont {Perlmutter}\ \emph {et~al.}(1999)\citenamefont {Perlmutter} \emph {et~al.}}]{SupernovaCosmologyProject:1998vns}%
  \BibitemOpen
  \bibfield  {author} {\bibinfo {author} {\bibfnamefont {S.}~\bibnamefont {Perlmutter}} \emph {et~al.} (\bibinfo {collaboration} {Supernova Cosmology Project}),\ }\bibfield  {title} {\bibinfo {title} {{Measurements of $\Omega$ and $\Lambda$ from 42 High Redshift Supernovae}},\ }\href {https://doi.org/10.1086/307221} {\bibfield  {journal} {\bibinfo  {journal} {Astrophys. J.}\ }\textbf {\bibinfo {volume} {517}},\ \bibinfo {pages} {565} (\bibinfo {year} {1999})},\ \Eprint {https://arxiv.org/abs/astro-ph/9812133} {arXiv:astro-ph/9812133} \BibitemShut {NoStop}%
\bibitem [{\citenamefont {Ade}\ \emph {et~al.}(2016)\citenamefont {Ade} \emph {et~al.}}]{Planck:2015fie}%
  \BibitemOpen
  \bibfield  {author} {\bibinfo {author} {\bibfnamefont {P.~A.~R.}\ \bibnamefont {Ade}} \emph {et~al.} (\bibinfo {collaboration} {Planck}),\ }\bibfield  {title} {\bibinfo {title} {{Planck 2015 results. XIII. Cosmological parameters}},\ }\href {https://doi.org/10.1051/0004-6361/201525830} {\bibfield  {journal} {\bibinfo  {journal} {Astron. Astrophys.}\ }\textbf {\bibinfo {volume} {594}},\ \bibinfo {pages} {A13} (\bibinfo {year} {2016})},\ \Eprint {https://arxiv.org/abs/1502.01589} {arXiv:1502.01589 [astro-ph.CO]} \BibitemShut {NoStop}%
\bibitem [{\citenamefont {Aghanim}\ \emph {et~al.}(2020{\natexlab{a}})\citenamefont {Aghanim} \emph {et~al.}}]{Planck:2018vyg}%
  \BibitemOpen
  \bibfield  {author} {\bibinfo {author} {\bibfnamefont {N.}~\bibnamefont {Aghanim}} \emph {et~al.} (\bibinfo {collaboration} {Planck}),\ }\bibfield  {title} {\bibinfo {title} {{Planck 2018 results. VI. Cosmological parameters}},\ }\href {https://doi.org/10.1051/0004-6361/201833910} {\bibfield  {journal} {\bibinfo  {journal} {Astron. Astrophys.}\ }\textbf {\bibinfo {volume} {641}},\ \bibinfo {pages} {A6} (\bibinfo {year} {2020}{\natexlab{a}})},\ \bibinfo {note} {[Erratum: Astron.Astrophys. 652, C4 (2021)]},\ \Eprint {https://arxiv.org/abs/1807.06209} {arXiv:1807.06209 [astro-ph.CO]} \BibitemShut {NoStop}%
\bibitem [{\citenamefont {Aghanim}\ \emph {et~al.}(2020{\natexlab{b}})\citenamefont {Aghanim} \emph {et~al.}}]{Planck:2019nip}%
  \BibitemOpen
  \bibfield  {author} {\bibinfo {author} {\bibfnamefont {N.}~\bibnamefont {Aghanim}} \emph {et~al.} (\bibinfo {collaboration} {Planck}),\ }\bibfield  {title} {\bibinfo {title} {{Planck 2018 results. V. CMB power spectra and likelihoods}},\ }\href {https://doi.org/10.1051/0004-6361/201936386} {\bibfield  {journal} {\bibinfo  {journal} {Astron. Astrophys.}\ }\textbf {\bibinfo {volume} {641}},\ \bibinfo {pages} {A5} (\bibinfo {year} {2020}{\natexlab{b}})},\ \Eprint {https://arxiv.org/abs/1907.12875} {arXiv:1907.12875 [astro-ph.CO]} \BibitemShut {NoStop}%
\bibitem [{\citenamefont {Scolnic}\ \emph {et~al.}(2018)\citenamefont {Scolnic} \emph {et~al.}}]{Pan-STARRS1:2017jku}%
  \BibitemOpen
  \bibfield  {author} {\bibinfo {author} {\bibfnamefont {D.~M.}\ \bibnamefont {Scolnic}} \emph {et~al.} (\bibinfo {collaboration} {Pan-STARRS1}),\ }\bibfield  {title} {\bibinfo {title} {{The Complete Light-curve Sample of Spectroscopically Confirmed SNe Ia from Pan-STARRS1 and Cosmological Constraints from the Combined Pantheon Sample}},\ }\href {https://doi.org/10.3847/1538-4357/aab9bb} {\bibfield  {journal} {\bibinfo  {journal} {Astrophys. J.}\ }\textbf {\bibinfo {volume} {859}},\ \bibinfo {pages} {101} (\bibinfo {year} {2018})},\ \Eprint {https://arxiv.org/abs/1710.00845} {arXiv:1710.00845 [astro-ph.CO]} \BibitemShut {NoStop}%
\bibitem [{\citenamefont {Abbott}\ \emph {et~al.}(2018)\citenamefont {Abbott} \emph {et~al.}}]{DES:2017myr}%
  \BibitemOpen
  \bibfield  {author} {\bibinfo {author} {\bibfnamefont {T.~M.~C.}\ \bibnamefont {Abbott}} \emph {et~al.} (\bibinfo {collaboration} {DES}),\ }\bibfield  {title} {\bibinfo {title} {{Dark Energy Survey year 1 results: Cosmological constraints from galaxy clustering and weak lensing}},\ }\href {https://doi.org/10.1103/PhysRevD.98.043526} {\bibfield  {journal} {\bibinfo  {journal} {Phys. Rev. D}\ }\textbf {\bibinfo {volume} {98}},\ \bibinfo {pages} {043526} (\bibinfo {year} {2018})},\ \Eprint {https://arxiv.org/abs/1708.01530} {arXiv:1708.01530 [astro-ph.CO]} \BibitemShut {NoStop}%
\bibitem [{\citenamefont {Alam}\ \emph {et~al.}(2021)\citenamefont {Alam} \emph {et~al.}}]{eBOSS:2020yzd}%
  \BibitemOpen
  \bibfield  {author} {\bibinfo {author} {\bibfnamefont {S.}~\bibnamefont {Alam}} \emph {et~al.} (\bibinfo {collaboration} {eBOSS}),\ }\bibfield  {title} {\bibinfo {title} {{Completed SDSS-IV extended Baryon Oscillation Spectroscopic Survey: Cosmological implications from two decades of spectroscopic surveys at the Apache Point Observatory}},\ }\href {https://doi.org/10.1103/PhysRevD.103.083533} {\bibfield  {journal} {\bibinfo  {journal} {Phys. Rev. D}\ }\textbf {\bibinfo {volume} {103}},\ \bibinfo {pages} {083533} (\bibinfo {year} {2021})},\ \Eprint {https://arxiv.org/abs/2007.08991} {arXiv:2007.08991 [astro-ph.CO]} \BibitemShut {NoStop}%
\bibitem [{\citenamefont {Klypin}\ \emph {et~al.}(1999)\citenamefont {Klypin}, \citenamefont {Kravtsov}, \citenamefont {Valenzuela},\ and\ \citenamefont {Prada}}]{Klypin:1999uc}%
  \BibitemOpen
  \bibfield  {author} {\bibinfo {author} {\bibfnamefont {A.~A.}\ \bibnamefont {Klypin}}, \bibinfo {author} {\bibfnamefont {A.~V.}\ \bibnamefont {Kravtsov}}, \bibinfo {author} {\bibfnamefont {O.}~\bibnamefont {Valenzuela}},\ and\ \bibinfo {author} {\bibfnamefont {F.}~\bibnamefont {Prada}},\ }\bibfield  {title} {\bibinfo {title} {{Where are the missing Galactic satellites?}},\ }\href {https://doi.org/10.1086/307643} {\bibfield  {journal} {\bibinfo  {journal} {Astrophys. J.}\ }\textbf {\bibinfo {volume} {522}},\ \bibinfo {pages} {82} (\bibinfo {year} {1999})},\ \Eprint {https://arxiv.org/abs/astro-ph/9901240} {arXiv:astro-ph/9901240} \BibitemShut {NoStop}%
\bibitem [{\citenamefont {Koposov}\ \emph {et~al.}(2008)\citenamefont {Koposov} \emph {et~al.}}]{Koposov:2007ni}%
  \BibitemOpen
  \bibfield  {author} {\bibinfo {author} {\bibfnamefont {S.}~\bibnamefont {Koposov}} \emph {et~al.},\ }\bibfield  {title} {\bibinfo {title} {{The Luminosity Function of the Milky Way Satellites}},\ }\href {https://doi.org/10.1086/589911} {\bibfield  {journal} {\bibinfo  {journal} {Astrophys. J.}\ }\textbf {\bibinfo {volume} {686}},\ \bibinfo {pages} {279} (\bibinfo {year} {2008})},\ \Eprint {https://arxiv.org/abs/0706.2687} {arXiv:0706.2687 [astro-ph]} \BibitemShut {NoStop}%
\bibitem [{\citenamefont {Boyarsky}\ \emph {et~al.}(2014)\citenamefont {Boyarsky}, \citenamefont {Ruchayskiy}, \citenamefont {Iakubovskyi},\ and\ \citenamefont {Franse}}]{Boyarsky:2014jta}%
  \BibitemOpen
  \bibfield  {author} {\bibinfo {author} {\bibfnamefont {A.}~\bibnamefont {Boyarsky}}, \bibinfo {author} {\bibfnamefont {O.}~\bibnamefont {Ruchayskiy}}, \bibinfo {author} {\bibfnamefont {D.}~\bibnamefont {Iakubovskyi}},\ and\ \bibinfo {author} {\bibfnamefont {J.}~\bibnamefont {Franse}},\ }\bibfield  {title} {\bibinfo {title} {{Unidentified Line in X-Ray Spectra of the Andromeda Galaxy and Perseus Galaxy Cluster}},\ }\href {https://doi.org/10.1103/PhysRevLett.113.251301} {\bibfield  {journal} {\bibinfo  {journal} {Phys. Rev. Lett.}\ }\textbf {\bibinfo {volume} {113}},\ \bibinfo {pages} {251301} (\bibinfo {year} {2014})},\ \Eprint {https://arxiv.org/abs/1402.4119} {arXiv:1402.4119 [astro-ph.CO]} \BibitemShut {NoStop}%
\bibitem [{\citenamefont {Bulbul}\ \emph {et~al.}(2014)\citenamefont {Bulbul}, \citenamefont {Markevitch}, \citenamefont {Foster}, \citenamefont {Smith}, \citenamefont {Loewenstein},\ and\ \citenamefont {Randall}}]{Bulbul:2014sua}%
  \BibitemOpen
  \bibfield  {author} {\bibinfo {author} {\bibfnamefont {E.}~\bibnamefont {Bulbul}}, \bibinfo {author} {\bibfnamefont {M.}~\bibnamefont {Markevitch}}, \bibinfo {author} {\bibfnamefont {A.}~\bibnamefont {Foster}}, \bibinfo {author} {\bibfnamefont {R.~K.}\ \bibnamefont {Smith}}, \bibinfo {author} {\bibfnamefont {M.}~\bibnamefont {Loewenstein}},\ and\ \bibinfo {author} {\bibfnamefont {S.~W.}\ \bibnamefont {Randall}},\ }\bibfield  {title} {\bibinfo {title} {{Detection of An Unidentified Emission Line in the Stacked X-ray spectrum of Galaxy Clusters}},\ }\href {https://doi.org/10.1088/0004-637X/789/1/13} {\bibfield  {journal} {\bibinfo  {journal} {Astrophys. J.}\ }\textbf {\bibinfo {volume} {789}},\ \bibinfo {pages} {13} (\bibinfo {year} {2014})},\ \Eprint {https://arxiv.org/abs/1402.2301} {arXiv:1402.2301 [astro-ph.CO]} \BibitemShut {NoStop}%
\bibitem [{\citenamefont {Spergel}\ and\ \citenamefont {Steinhardt}(2000)}]{Spergel:1999mh}%
  \BibitemOpen
  \bibfield  {author} {\bibinfo {author} {\bibfnamefont {D.~N.}\ \bibnamefont {Spergel}}\ and\ \bibinfo {author} {\bibfnamefont {P.~J.}\ \bibnamefont {Steinhardt}},\ }\bibfield  {title} {\bibinfo {title} {{Observational evidence for selfinteracting cold dark matter}},\ }\href {https://doi.org/10.1103/PhysRevLett.84.3760} {\bibfield  {journal} {\bibinfo  {journal} {Phys. Rev. Lett.}\ }\textbf {\bibinfo {volume} {84}},\ \bibinfo {pages} {3760} (\bibinfo {year} {2000})},\ \Eprint {https://arxiv.org/abs/astro-ph/9909386} {arXiv:astro-ph/9909386} \BibitemShut {NoStop}%
\bibitem [{\citenamefont {Colombi}\ \emph {et~al.}(1996)\citenamefont {Colombi}, \citenamefont {Dodelson},\ and\ \citenamefont {Widrow}}]{Colombi:1995ze}%
  \BibitemOpen
  \bibfield  {author} {\bibinfo {author} {\bibfnamefont {S.}~\bibnamefont {Colombi}}, \bibinfo {author} {\bibfnamefont {S.}~\bibnamefont {Dodelson}},\ and\ \bibinfo {author} {\bibfnamefont {L.~M.}\ \bibnamefont {Widrow}},\ }\bibfield  {title} {\bibinfo {title} {{Large scale structure tests of warm dark matter}},\ }\href {https://doi.org/10.1086/176788} {\bibfield  {journal} {\bibinfo  {journal} {Astrophys. J.}\ }\textbf {\bibinfo {volume} {458}},\ \bibinfo {pages} {1} (\bibinfo {year} {1996})},\ \Eprint {https://arxiv.org/abs/astro-ph/9505029} {arXiv:astro-ph/9505029} \BibitemShut {NoStop}%
\bibitem [{\citenamefont {Marsh}(2016)}]{Marsh:2015xka}%
  \BibitemOpen
  \bibfield  {author} {\bibinfo {author} {\bibfnamefont {D.~J.~E.}\ \bibnamefont {Marsh}},\ }\bibfield  {title} {\bibinfo {title} {{Axion Cosmology}},\ }\href {https://doi.org/10.1016/j.physrep.2016.06.005} {\bibfield  {journal} {\bibinfo  {journal} {Phys. Rept.}\ }\textbf {\bibinfo {volume} {643}},\ \bibinfo {pages} {1} (\bibinfo {year} {2016})},\ \Eprint {https://arxiv.org/abs/1510.07633} {arXiv:1510.07633 [astro-ph.CO]} \BibitemShut {NoStop}%
\bibitem [{\citenamefont {O'Hare}(2024)}]{OHare:2024nmr}%
  \BibitemOpen
  \bibfield  {author} {\bibinfo {author} {\bibfnamefont {C.~A.~J.}\ \bibnamefont {O'Hare}},\ }\bibfield  {title} {\bibinfo {title} {{Cosmology of axion dark matter}},\ }\href {https://doi.org/10.22323/1.454.0040} {\bibfield  {journal} {\bibinfo  {journal} {PoS}\ }\textbf {\bibinfo {volume} {COSMICWISPers}},\ \bibinfo {pages} {040} (\bibinfo {year} {2024})},\ \Eprint {https://arxiv.org/abs/2403.17697} {arXiv:2403.17697 [hep-ph]} \BibitemShut {NoStop}%
\bibitem [{\citenamefont {Hui}\ \emph {et~al.}(2017)\citenamefont {Hui}, \citenamefont {Ostriker}, \citenamefont {Tremaine},\ and\ \citenamefont {Witten}}]{Hui:2016ltb}%
  \BibitemOpen
  \bibfield  {author} {\bibinfo {author} {\bibfnamefont {L.}~\bibnamefont {Hui}}, \bibinfo {author} {\bibfnamefont {J.~P.}\ \bibnamefont {Ostriker}}, \bibinfo {author} {\bibfnamefont {S.}~\bibnamefont {Tremaine}},\ and\ \bibinfo {author} {\bibfnamefont {E.}~\bibnamefont {Witten}},\ }\bibfield  {title} {\bibinfo {title} {{Ultralight scalars as cosmological dark matter}},\ }\href {https://doi.org/10.1103/PhysRevD.95.043541} {\bibfield  {journal} {\bibinfo  {journal} {Phys. Rev. D}\ }\textbf {\bibinfo {volume} {95}},\ \bibinfo {pages} {043541} (\bibinfo {year} {2017})},\ \Eprint {https://arxiv.org/abs/1610.08297} {arXiv:1610.08297 [astro-ph.CO]} \BibitemShut {NoStop}%
\bibitem [{\citenamefont {Schneider}\ \emph {et~al.}(1992)\citenamefont {Schneider}, \citenamefont {Ehlers},\ and\ \citenamefont {Falco}}]{Schneider:1992bmb}%
  \BibitemOpen
  \bibfield  {author} {\bibinfo {author} {\bibfnamefont {P.}~\bibnamefont {Schneider}}, \bibinfo {author} {\bibfnamefont {J.}~\bibnamefont {Ehlers}},\ and\ \bibinfo {author} {\bibfnamefont {E.~E.}\ \bibnamefont {Falco}},\ }\href {https://doi.org/10.1007/978-3-662-03758-4} {\emph {\bibinfo {title} {{Gravitational Lenses}}}},\ Astronomy and Astrophysics Library\ (\bibinfo  {publisher} {Springer},\ \bibinfo {year} {1992})\BibitemShut {NoStop}%
\bibitem [{\citenamefont {Collett}\ \emph {et~al.}(2017)\citenamefont {Collett} \emph {et~al.}}]{DES:2017nhh}%
  \BibitemOpen
  \bibfield  {author} {\bibinfo {author} {\bibfnamefont {T.~E.}\ \bibnamefont {Collett}} \emph {et~al.} (\bibinfo {collaboration} {DES}),\ }\bibfield  {title} {\bibinfo {title} {{Core or Cusps: The Central Dark Matter Profile of a Strong Lensing Cluster with a Bright Central Image at Redshift 1}},\ }\href {https://doi.org/10.3847/1538-4357/aa76e6} {\bibfield  {journal} {\bibinfo  {journal} {Astrophys. J.}\ }\textbf {\bibinfo {volume} {843}},\ \bibinfo {pages} {148} (\bibinfo {year} {2017})},\ \Eprint {https://arxiv.org/abs/1703.08410} {arXiv:1703.08410 [astro-ph.CO]} \BibitemShut {NoStop}%
\bibitem [{\citenamefont {Mahler}\ \emph {et~al.}(2023)\citenamefont {Mahler} \emph {et~al.}}]{Mahler:2022tgu}%
  \BibitemOpen
  \bibfield  {author} {\bibinfo {author} {\bibfnamefont {G.}~\bibnamefont {Mahler}} \emph {et~al.},\ }\bibfield  {title} {\bibinfo {title} {{Precision Modeling of JWST's First Cluster Lens SMACS J0723.3\textendash{}7327*}},\ }\href {https://doi.org/10.3847/1538-4357/acaea9} {\bibfield  {journal} {\bibinfo  {journal} {Astrophys. J.}\ }\textbf {\bibinfo {volume} {945}},\ \bibinfo {pages} {49} (\bibinfo {year} {2023})},\ \Eprint {https://arxiv.org/abs/2207.07101} {arXiv:2207.07101 [astro-ph.GA]} \BibitemShut {NoStop}%
\bibitem [{\citenamefont {Bartelmann}(2010)}]{Bartelmann:2010fz}%
  \BibitemOpen
  \bibfield  {author} {\bibinfo {author} {\bibfnamefont {M.}~\bibnamefont {Bartelmann}},\ }\bibfield  {title} {\bibinfo {title} {{Gravitational Lensing}},\ }\href {https://doi.org/10.1088/0264-9381/27/23/233001} {\bibfield  {journal} {\bibinfo  {journal} {Class. Quant. Grav.}\ }\textbf {\bibinfo {volume} {27}},\ \bibinfo {pages} {233001} (\bibinfo {year} {2010})},\ \Eprint {https://arxiv.org/abs/1010.3829} {arXiv:1010.3829 [astro-ph.CO]} \BibitemShut {NoStop}%
\bibitem [{\citenamefont {Shapiro}\ \emph {et~al.}(2004)\citenamefont {Shapiro}, \citenamefont {Davis}, \citenamefont {Lebach},\ and\ \citenamefont {Gregory}}]{Shapiro:2004zz}%
  \BibitemOpen
  \bibfield  {author} {\bibinfo {author} {\bibfnamefont {S.~S.}\ \bibnamefont {Shapiro}}, \bibinfo {author} {\bibfnamefont {J.~L.}\ \bibnamefont {Davis}}, \bibinfo {author} {\bibfnamefont {D.~E.}\ \bibnamefont {Lebach}},\ and\ \bibinfo {author} {\bibfnamefont {J.~S.}\ \bibnamefont {Gregory}},\ }\bibfield  {title} {\bibinfo {title} {{Measurement of the Solar Gravitational Deflection of Radio Waves using Geodetic Very-Long-Baseline Interferometry Data, 1979-1999}},\ }\href {https://doi.org/10.1103/PhysRevLett.92.121101} {\bibfield  {journal} {\bibinfo  {journal} {Phys. Rev. Lett.}\ }\textbf {\bibinfo {volume} {92}},\ \bibinfo {pages} {121101} (\bibinfo {year} {2004})}\BibitemShut {NoStop}%
\bibitem [{\citenamefont {Kelly}\ \emph {et~al.}(2023)\citenamefont {Kelly} \emph {et~al.}}]{Kelly:2023mgv}%
  \BibitemOpen
  \bibfield  {author} {\bibinfo {author} {\bibfnamefont {P.~L.}\ \bibnamefont {Kelly}} \emph {et~al.},\ }\bibfield  {title} {\bibinfo {title} {{Constraints on the Hubble constant from supernova Refsdal\textquoteright{}s reappearance}},\ }\href {https://doi.org/10.1126/science.abh1322} {\bibfield  {journal} {\bibinfo  {journal} {Science}\ }\textbf {\bibinfo {volume} {380}},\ \bibinfo {pages} {abh1322} (\bibinfo {year} {2023})},\ \Eprint {https://arxiv.org/abs/2305.06367} {arXiv:2305.06367 [astro-ph.CO]} \BibitemShut {NoStop}%
\bibitem [{\citenamefont {Pascale}\ \emph {et~al.}(2024)\citenamefont {Pascale} \emph {et~al.}}]{Pascale:2024qjr}%
  \BibitemOpen
  \bibfield  {author} {\bibinfo {author} {\bibfnamefont {M.}~\bibnamefont {Pascale}} \emph {et~al.},\ }\bibfield  {title} {\bibinfo {title} {{SN H0pe: The First Measurement of $H_0$ from a Multiply-Imaged Type Ia Supernova, Discovered by JWST}},\ }\href@noop {} {\  (\bibinfo {year} {2024})},\ \Eprint {https://arxiv.org/abs/2403.18902} {arXiv:2403.18902 [astro-ph.CO]} \BibitemShut {NoStop}%
\bibitem [{\citenamefont {Niikura}\ \emph {et~al.}(2019)\citenamefont {Niikura} \emph {et~al.}}]{Niikura:2017zjd}%
  \BibitemOpen
  \bibfield  {author} {\bibinfo {author} {\bibfnamefont {H.}~\bibnamefont {Niikura}} \emph {et~al.},\ }\bibfield  {title} {\bibinfo {title} {{Microlensing constraints on primordial black holes with Subaru/HSC Andromeda observations}},\ }\href {https://doi.org/10.1038/s41550-019-0723-1} {\bibfield  {journal} {\bibinfo  {journal} {Nature Astron.}\ }\textbf {\bibinfo {volume} {3}},\ \bibinfo {pages} {524} (\bibinfo {year} {2019})},\ \Eprint {https://arxiv.org/abs/1701.02151} {arXiv:1701.02151 [astro-ph.CO]} \BibitemShut {NoStop}%
\bibitem [{\citenamefont {Abbott}\ \emph {et~al.}(2021)\citenamefont {Abbott} \emph {et~al.}}]{LIGOScientific:2021izm}%
  \BibitemOpen
  \bibfield  {author} {\bibinfo {author} {\bibfnamefont {R.}~\bibnamefont {Abbott}} \emph {et~al.} (\bibinfo {collaboration} {LIGO Scientific, VIRGO}),\ }\bibfield  {title} {\bibinfo {title} {{Search for Lensing Signatures in the Gravitational-Wave Observations from the First Half of LIGO\textendash{}Virgo\textquoteright{}s Third Observing Run}},\ }\href {https://doi.org/10.3847/1538-4357/ac23db} {\bibfield  {journal} {\bibinfo  {journal} {Astrophys. J.}\ }\textbf {\bibinfo {volume} {923}},\ \bibinfo {pages} {14} (\bibinfo {year} {2021})},\ \Eprint {https://arxiv.org/abs/2105.06384} {arXiv:2105.06384 [gr-qc]} \BibitemShut {NoStop}%
\bibitem [{\citenamefont {Abbott}\ \emph {et~al.}(2024)\citenamefont {Abbott} \emph {et~al.}}]{LIGOScientific:2023bwz}%
  \BibitemOpen
  \bibfield  {author} {\bibinfo {author} {\bibfnamefont {R.}~\bibnamefont {Abbott}} \emph {et~al.} (\bibinfo {collaboration} {LIGO Scientific, KAGRA, VIRGO}),\ }\bibfield  {title} {\bibinfo {title} {{Search for Gravitational-lensing Signatures in the Full Third Observing Run of the LIGO\textendash{}Virgo Network}},\ }\href {https://doi.org/10.3847/1538-4357/ad3e83} {\bibfield  {journal} {\bibinfo  {journal} {Astrophys. J.}\ }\textbf {\bibinfo {volume} {970}},\ \bibinfo {pages} {191} (\bibinfo {year} {2024})},\ \Eprint {https://arxiv.org/abs/2304.08393} {arXiv:2304.08393 [gr-qc]} \BibitemShut {NoStop}%
\bibitem [{\citenamefont {Chan}\ \emph {et~al.}(2025)\citenamefont {Chan}, \citenamefont {Seo}, \citenamefont {Li}, \citenamefont {Fong},\ and\ \citenamefont {Ezquiaga}}]{Chan:2024qmb}%
  \BibitemOpen
  \bibfield  {author} {\bibinfo {author} {\bibfnamefont {J.~C.~L.}\ \bibnamefont {Chan}}, \bibinfo {author} {\bibfnamefont {E.}~\bibnamefont {Seo}}, \bibinfo {author} {\bibfnamefont {A.~K.~Y.}\ \bibnamefont {Li}}, \bibinfo {author} {\bibfnamefont {H.}~\bibnamefont {Fong}},\ and\ \bibinfo {author} {\bibfnamefont {J.~M.}\ \bibnamefont {Ezquiaga}},\ }\bibfield  {title} {\bibinfo {title} {{Detectability of lensed gravitational waves in matched-filtering searches}},\ }\href {https://doi.org/10.1103/PhysRevD.111.084019} {\bibfield  {journal} {\bibinfo  {journal} {Phys. Rev. D}\ }\textbf {\bibinfo {volume} {111}},\ \bibinfo {pages} {084019} (\bibinfo {year} {2025})},\ \Eprint {https://arxiv.org/abs/2411.13058} {arXiv:2411.13058 [gr-qc]} \BibitemShut {NoStop}%
\bibitem [{\citenamefont {Jung}\ and\ \citenamefont {Shin}(2019)}]{Jung:2017flg}%
  \BibitemOpen
  \bibfield  {author} {\bibinfo {author} {\bibfnamefont {S.}~\bibnamefont {Jung}}\ and\ \bibinfo {author} {\bibfnamefont {C.~S.}\ \bibnamefont {Shin}},\ }\bibfield  {title} {\bibinfo {title} {{Gravitational-Wave Fringes at LIGO: Detecting Compact Dark Matter by Gravitational Lensing}},\ }\href {https://doi.org/10.1103/PhysRevLett.122.041103} {\bibfield  {journal} {\bibinfo  {journal} {Phys. Rev. Lett.}\ }\textbf {\bibinfo {volume} {122}},\ \bibinfo {pages} {041103} (\bibinfo {year} {2019})},\ \Eprint {https://arxiv.org/abs/1712.01396} {arXiv:1712.01396 [astro-ph.CO]} \BibitemShut {NoStop}%
\bibitem [{\citenamefont {Urrutia}\ and\ \citenamefont {Vaskonen}(2021)}]{Urrutia:2021qak}%
  \BibitemOpen
  \bibfield  {author} {\bibinfo {author} {\bibfnamefont {J.}~\bibnamefont {Urrutia}}\ and\ \bibinfo {author} {\bibfnamefont {V.}~\bibnamefont {Vaskonen}},\ }\bibfield  {title} {\bibinfo {title} {{Lensing of gravitational waves as a probe of compact dark matter}},\ }\href {https://doi.org/10.1093/mnras/stab3118} {\bibfield  {journal} {\bibinfo  {journal} {Mon. Not. Roy. Astron. Soc.}\ }\textbf {\bibinfo {volume} {509}},\ \bibinfo {pages} {1358} (\bibinfo {year} {2021})},\ \Eprint {https://arxiv.org/abs/2109.03213} {arXiv:2109.03213 [astro-ph.CO]} \BibitemShut {NoStop}%
\bibitem [{\citenamefont {Oguri}\ and\ \citenamefont {Takahashi}(2020)}]{Oguri:2020ldf}%
  \BibitemOpen
  \bibfield  {author} {\bibinfo {author} {\bibfnamefont {M.}~\bibnamefont {Oguri}}\ and\ \bibinfo {author} {\bibfnamefont {R.}~\bibnamefont {Takahashi}},\ }\bibfield  {title} {\bibinfo {title} {{Probing Dark Low-mass Halos and Primordial Black Holes with Frequency-dependent Gravitational Lensing Dispersions of Gravitational Waves}},\ }\href {https://doi.org/10.3847/1538-4357/abafab} {\bibfield  {journal} {\bibinfo  {journal} {Astrophys. J.}\ }\textbf {\bibinfo {volume} {901}},\ \bibinfo {pages} {58} (\bibinfo {year} {2020})},\ \Eprint {https://arxiv.org/abs/2007.01936} {arXiv:2007.01936 [astro-ph.CO]} \BibitemShut {NoStop}%
\bibitem [{\citenamefont {Oguri}\ and\ \citenamefont {Takahashi}(2022)}]{Oguri:2022zpn}%
  \BibitemOpen
  \bibfield  {author} {\bibinfo {author} {\bibfnamefont {M.}~\bibnamefont {Oguri}}\ and\ \bibinfo {author} {\bibfnamefont {R.}~\bibnamefont {Takahashi}},\ }\bibfield  {title} {\bibinfo {title} {{Amplitude and phase fluctuations of gravitational waves magnified by strong gravitational lensing}},\ }\href {https://doi.org/10.1103/PhysRevD.106.043532} {\bibfield  {journal} {\bibinfo  {journal} {Phys. Rev. D}\ }\textbf {\bibinfo {volume} {106}},\ \bibinfo {pages} {043532} (\bibinfo {year} {2022})},\ \Eprint {https://arxiv.org/abs/2204.00814} {arXiv:2204.00814 [astro-ph.CO]} \BibitemShut {NoStop}%
\bibitem [{\citenamefont {Nakamura}(1998)}]{Nakamura:1997sw}%
  \BibitemOpen
  \bibfield  {author} {\bibinfo {author} {\bibfnamefont {T.~T.}\ \bibnamefont {Nakamura}},\ }\bibfield  {title} {\bibinfo {title} {{Gravitational lensing of gravitational waves from inspiraling binaries by a point mass lens}},\ }\href {https://doi.org/10.1103/PhysRevLett.80.1138} {\bibfield  {journal} {\bibinfo  {journal} {Phys. Rev. Lett.}\ }\textbf {\bibinfo {volume} {80}},\ \bibinfo {pages} {1138} (\bibinfo {year} {1998})}\BibitemShut {NoStop}%
\bibitem [{\citenamefont {Nakamura}\ and\ \citenamefont {Deguchi}(1999)}]{Nakamura:1999uwi}%
  \BibitemOpen
  \bibfield  {author} {\bibinfo {author} {\bibfnamefont {T.~T.}\ \bibnamefont {Nakamura}}\ and\ \bibinfo {author} {\bibfnamefont {S.}~\bibnamefont {Deguchi}},\ }\bibfield  {title} {\bibinfo {title} {{Wave Optics in Gravitational Lensing}},\ }\href {https://doi.org/10.1143/ptps.133.137} {\bibfield  {journal} {\bibinfo  {journal} {Prog. Theor. Phys. Suppl.}\ }\textbf {\bibinfo {volume} {133}},\ \bibinfo {pages} {137} (\bibinfo {year} {1999})}\BibitemShut {NoStop}%
\bibitem [{\citenamefont {Takahashi}\ and\ \citenamefont {Nakamura}(2003)}]{Takahashi:2003ix}%
  \BibitemOpen
  \bibfield  {author} {\bibinfo {author} {\bibfnamefont {R.}~\bibnamefont {Takahashi}}\ and\ \bibinfo {author} {\bibfnamefont {T.}~\bibnamefont {Nakamura}},\ }\bibfield  {title} {\bibinfo {title} {{Wave effects in gravitational lensing of gravitational waves from chirping binaries}},\ }\href {https://doi.org/10.1086/377430} {\bibfield  {journal} {\bibinfo  {journal} {Astrophys. J.}\ }\textbf {\bibinfo {volume} {595}},\ \bibinfo {pages} {1039} (\bibinfo {year} {2003})},\ \Eprint {https://arxiv.org/abs/astro-ph/0305055} {arXiv:astro-ph/0305055} \BibitemShut {NoStop}%
\bibitem [{\citenamefont {Cheung}\ \emph {et~al.}(2024)\citenamefont {Cheung}, \citenamefont {Ng}, \citenamefont {Zumalac\'arregui},\ and\ \citenamefont {Berti}}]{Cheung:2024ugg}%
  \BibitemOpen
  \bibfield  {author} {\bibinfo {author} {\bibfnamefont {M.~H.-Y.}\ \bibnamefont {Cheung}}, \bibinfo {author} {\bibfnamefont {K.~K.~Y.}\ \bibnamefont {Ng}}, \bibinfo {author} {\bibfnamefont {M.}~\bibnamefont {Zumalac\'arregui}},\ and\ \bibinfo {author} {\bibfnamefont {E.}~\bibnamefont {Berti}},\ }\bibfield  {title} {\bibinfo {title} {{Probing minihalo lenses with diffracted gravitational waves}},\ }\href {https://doi.org/10.1103/PhysRevD.109.124020} {\bibfield  {journal} {\bibinfo  {journal} {Phys. Rev. D}\ }\textbf {\bibinfo {volume} {109}},\ \bibinfo {pages} {124020} (\bibinfo {year} {2024})},\ \Eprint {https://arxiv.org/abs/2403.13876} {arXiv:2403.13876 [gr-qc]} \BibitemShut {NoStop}%
\bibitem [{\citenamefont {Tambalo}\ \emph {et~al.}(2023)\citenamefont {Tambalo}, \citenamefont {Zumalac\'arregui}, \citenamefont {Dai},\ and\ \citenamefont {Cheung}}]{Tambalo:2022wlm}%
  \BibitemOpen
  \bibfield  {author} {\bibinfo {author} {\bibfnamefont {G.}~\bibnamefont {Tambalo}}, \bibinfo {author} {\bibfnamefont {M.}~\bibnamefont {Zumalac\'arregui}}, \bibinfo {author} {\bibfnamefont {L.}~\bibnamefont {Dai}},\ and\ \bibinfo {author} {\bibfnamefont {M.~H.-Y.}\ \bibnamefont {Cheung}},\ }\bibfield  {title} {\bibinfo {title} {{Gravitational wave lensing as a probe of halo properties and dark matter}},\ }\href {https://doi.org/10.1103/PhysRevD.108.103529} {\bibfield  {journal} {\bibinfo  {journal} {Phys. Rev. D}\ }\textbf {\bibinfo {volume} {108}},\ \bibinfo {pages} {103529} (\bibinfo {year} {2023})},\ \Eprint {https://arxiv.org/abs/2212.11960} {arXiv:2212.11960 [astro-ph.CO]} \BibitemShut {NoStop}%
\bibitem [{\citenamefont {Sesana}(2016)}]{Sesana:2016ljz}%
  \BibitemOpen
  \bibfield  {author} {\bibinfo {author} {\bibfnamefont {A.}~\bibnamefont {Sesana}},\ }\bibfield  {title} {\bibinfo {title} {{Prospects for Multiband Gravitational-Wave Astronomy after GW150914}},\ }\href {https://doi.org/10.1103/PhysRevLett.116.231102} {\bibfield  {journal} {\bibinfo  {journal} {Phys. Rev. Lett.}\ }\textbf {\bibinfo {volume} {116}},\ \bibinfo {pages} {231102} (\bibinfo {year} {2016})},\ \Eprint {https://arxiv.org/abs/1602.06951} {arXiv:1602.06951 [gr-qc]} \BibitemShut {NoStop}%
\bibitem [{\citenamefont {Muttoni}\ \emph {et~al.}(2022)\citenamefont {Muttoni}, \citenamefont {Mangiagli}, \citenamefont {Sesana}, \citenamefont {Laghi}, \citenamefont {Del~Pozzo}, \citenamefont {Izquierdo-Villalba},\ and\ \citenamefont {Rosati}}]{Muttoni:2021veo}%
  \BibitemOpen
  \bibfield  {author} {\bibinfo {author} {\bibfnamefont {N.}~\bibnamefont {Muttoni}}, \bibinfo {author} {\bibfnamefont {A.}~\bibnamefont {Mangiagli}}, \bibinfo {author} {\bibfnamefont {A.}~\bibnamefont {Sesana}}, \bibinfo {author} {\bibfnamefont {D.}~\bibnamefont {Laghi}}, \bibinfo {author} {\bibfnamefont {W.}~\bibnamefont {Del~Pozzo}}, \bibinfo {author} {\bibfnamefont {D.}~\bibnamefont {Izquierdo-Villalba}},\ and\ \bibinfo {author} {\bibfnamefont {M.}~\bibnamefont {Rosati}},\ }\bibfield  {title} {\bibinfo {title} {{Multiband gravitational wave cosmology with stellar origin black hole binaries}},\ }\href {https://doi.org/10.1103/PhysRevD.105.043509} {\bibfield  {journal} {\bibinfo  {journal} {Phys. Rev. D}\ }\textbf {\bibinfo {volume} {105}},\ \bibinfo {pages} {043509} (\bibinfo {year} {2022})},\ \Eprint {https://arxiv.org/abs/2109.13934} {arXiv:2109.13934 [astro-ph.CO]} \BibitemShut {NoStop}%
\bibitem [{\citenamefont {Kawamura}\ \emph {et~al.}(2011)\citenamefont {Kawamura} \emph {et~al.}}]{Kawamura:2011zz}%
  \BibitemOpen
  \bibfield  {author} {\bibinfo {author} {\bibfnamefont {S.}~\bibnamefont {Kawamura}} \emph {et~al.},\ }\bibfield  {title} {\bibinfo {title} {{The Japanese space gravitational wave antenna: DECIGO}},\ }\href {https://doi.org/10.1088/0264-9381/28/9/094011} {\bibfield  {journal} {\bibinfo  {journal} {Class. Quant. Grav.}\ }\textbf {\bibinfo {volume} {28}},\ \bibinfo {pages} {094011} (\bibinfo {year} {2011})}\BibitemShut {NoStop}%
\bibitem [{\citenamefont {Punturo}\ \emph {et~al.}(2010)\citenamefont {Punturo} \emph {et~al.}}]{Punturo:2010zz}%
  \BibitemOpen
  \bibfield  {author} {\bibinfo {author} {\bibfnamefont {M.}~\bibnamefont {Punturo}} \emph {et~al.},\ }\bibfield  {title} {\bibinfo {title} {{The Einstein Telescope: A third-generation gravitational wave observatory}},\ }\href {https://doi.org/10.1088/0264-9381/27/19/194002} {\bibfield  {journal} {\bibinfo  {journal} {Class. Quant. Grav.}\ }\textbf {\bibinfo {volume} {27}},\ \bibinfo {pages} {194002} (\bibinfo {year} {2010})}\BibitemShut {NoStop}%
\bibitem [{\citenamefont {Kawamura}\ \emph {et~al.}(2019)\citenamefont {Kawamura} \emph {et~al.}}]{Kawamura:2018esd}%
  \BibitemOpen
  \bibfield  {author} {\bibinfo {author} {\bibfnamefont {S.}~\bibnamefont {Kawamura}} \emph {et~al.},\ }\bibfield  {title} {\bibinfo {title} {{Space gravitational-wave antennas DECIGO and B-DECIGO}},\ }\href {https://doi.org/10.1142/S0218271818450013} {\bibfield  {journal} {\bibinfo  {journal} {Int. J. Mod. Phys. D}\ }\textbf {\bibinfo {volume} {28}},\ \bibinfo {pages} {1845001} (\bibinfo {year} {2019})}\BibitemShut {NoStop}%
\bibitem [{\citenamefont {Grimm}\ and\ \citenamefont {Harms}(2020)}]{Grimm:2020ivq}%
  \BibitemOpen
  \bibfield  {author} {\bibinfo {author} {\bibfnamefont {S.}~\bibnamefont {Grimm}}\ and\ \bibinfo {author} {\bibfnamefont {J.}~\bibnamefont {Harms}},\ }\bibfield  {title} {\bibinfo {title} {{Multiband gravitational-wave parameter estimation: A study of future detectors}},\ }\href {https://doi.org/10.1103/PhysRevD.102.022007} {\bibfield  {journal} {\bibinfo  {journal} {Phys. Rev. D}\ }\textbf {\bibinfo {volume} {102}},\ \bibinfo {pages} {022007} (\bibinfo {year} {2020})},\ \Eprint {https://arxiv.org/abs/2004.01434} {arXiv:2004.01434 [gr-qc]} \BibitemShut {NoStop}%
\bibitem [{\citenamefont {Nakano}\ \emph {et~al.}(2021)\citenamefont {Nakano}, \citenamefont {Fujita}, \citenamefont {Isoyama},\ and\ \citenamefont {Sago}}]{Nakano:2021bbw}%
  \BibitemOpen
  \bibfield  {author} {\bibinfo {author} {\bibfnamefont {H.}~\bibnamefont {Nakano}}, \bibinfo {author} {\bibfnamefont {R.}~\bibnamefont {Fujita}}, \bibinfo {author} {\bibfnamefont {S.}~\bibnamefont {Isoyama}},\ and\ \bibinfo {author} {\bibfnamefont {N.}~\bibnamefont {Sago}},\ }\bibfield  {title} {\bibinfo {title} {{Scope out multiband gravitational-wave observations of GW190521-like binary black holes with space gravitational wave antenna B-DECIGO}},\ }\href {https://doi.org/10.3390/universe7030053} {\bibfield  {journal} {\bibinfo  {journal} {Universe}\ }\textbf {\bibinfo {volume} {7}},\ \bibinfo {pages} {53} (\bibinfo {year} {2021})},\ \Eprint {https://arxiv.org/abs/2101.06402} {arXiv:2101.06402 [gr-qc]} \BibitemShut {NoStop}%
\bibitem [{\citenamefont {Peters}(1974)}]{Peters:1974gj}%
  \BibitemOpen
  \bibfield  {author} {\bibinfo {author} {\bibfnamefont {P.~C.}\ \bibnamefont {Peters}},\ }\bibfield  {title} {\bibinfo {title} {{Index of refraction for scalar, electromagnetic, and gravitational waves in weak gravitational fields}},\ }\href {https://doi.org/10.1103/PhysRevD.9.2207} {\bibfield  {journal} {\bibinfo  {journal} {Phys. Rev. D}\ }\textbf {\bibinfo {volume} {9}},\ \bibinfo {pages} {2207} (\bibinfo {year} {1974})}\BibitemShut {NoStop}%
\bibitem [{\citenamefont {Poon}\ \emph {et~al.}(2024)\citenamefont {Poon}, \citenamefont {Rinaldi}, \citenamefont {Janquart}, \citenamefont {Narola},\ and\ \citenamefont {Hannuksela}}]{Poon:2024zxn}%
  \BibitemOpen
  \bibfield  {author} {\bibinfo {author} {\bibfnamefont {J.~S.~C.}\ \bibnamefont {Poon}}, \bibinfo {author} {\bibfnamefont {S.}~\bibnamefont {Rinaldi}}, \bibinfo {author} {\bibfnamefont {J.}~\bibnamefont {Janquart}}, \bibinfo {author} {\bibfnamefont {H.}~\bibnamefont {Narola}},\ and\ \bibinfo {author} {\bibfnamefont {O.~A.}\ \bibnamefont {Hannuksela}},\ }\bibfield  {title} {\bibinfo {title} {{Galaxy lens reconstruction based on strongly lensed gravitational waves: similarity transformation degeneracy and mass-sheet degeneracy}},\ }\href@noop {} {\  (\bibinfo {year} {2024})},\ \Eprint {https://arxiv.org/abs/2406.06463} {arXiv:2406.06463 [astro-ph.HE]} \BibitemShut {NoStop}%
\bibitem [{\citenamefont {Matsunaga}\ and\ \citenamefont {Yamamoto}(2006)}]{Matsunaga:2006uc}%
  \BibitemOpen
  \bibfield  {author} {\bibinfo {author} {\bibfnamefont {N.}~\bibnamefont {Matsunaga}}\ and\ \bibinfo {author} {\bibfnamefont {K.}~\bibnamefont {Yamamoto}},\ }\bibfield  {title} {\bibinfo {title} {{The finite source size effect and the wave optics in gravitational lensing}},\ }\href {https://doi.org/10.1088/1475-7516/2006/01/023} {\bibfield  {journal} {\bibinfo  {journal} {JCAP}\ }\textbf {\bibinfo {volume} {01}},\ \bibinfo {pages} {023}},\ \Eprint {https://arxiv.org/abs/astro-ph/0601701} {arXiv:astro-ph/0601701} \BibitemShut {NoStop}%
\bibitem [{\citenamefont {{Kormann}}\ \emph {et~al.}(1994)\citenamefont {{Kormann}}, \citenamefont {{Schneider}},\ and\ \citenamefont {{Bartelmann}}}]{1994A&A...284..285K}%
  \BibitemOpen
  \bibfield  {author} {\bibinfo {author} {\bibfnamefont {R.}~\bibnamefont {{Kormann}}}, \bibinfo {author} {\bibfnamefont {P.}~\bibnamefont {{Schneider}}},\ and\ \bibinfo {author} {\bibfnamefont {M.}~\bibnamefont {{Bartelmann}}},\ }\bibfield  {title} {\bibinfo {title} {{Isothermal elliptical gravitational lens models.}},\ }\href@noop {} {\bibfield  {journal} {\bibinfo  {journal} {Astronomy and Astrophysics}\ }\textbf {\bibinfo {volume} {284}},\ \bibinfo {pages} {285} (\bibinfo {year} {1994})}\BibitemShut {NoStop}%
\bibitem [{\citenamefont {Flores}\ and\ \citenamefont {Primack}(1996)}]{Flores:1995dc}%
  \BibitemOpen
  \bibfield  {author} {\bibinfo {author} {\bibfnamefont {R.~A.}\ \bibnamefont {Flores}}\ and\ \bibinfo {author} {\bibfnamefont {J.~R.}\ \bibnamefont {Primack}},\ }\bibfield  {title} {\bibinfo {title} {{Cluster cores, gravitational lensing, and cosmology}},\ }\href {https://doi.org/10.1086/309879} {\bibfield  {journal} {\bibinfo  {journal} {Astrophys. J. Lett.}\ }\textbf {\bibinfo {volume} {457}},\ \bibinfo {pages} {L5} (\bibinfo {year} {1996})},\ \Eprint {https://arxiv.org/abs/astro-ph/9512063} {arXiv:astro-ph/9512063} \BibitemShut {NoStop}%
\bibitem [{\citenamefont {Treu}(2010)}]{Treu:2010uj}%
  \BibitemOpen
  \bibfield  {author} {\bibinfo {author} {\bibfnamefont {T.}~\bibnamefont {Treu}},\ }\bibfield  {title} {\bibinfo {title} {{Strong Lensing by Galaxies}},\ }\href {https://doi.org/10.1146/annurev-astro-081309-130924} {\bibfield  {journal} {\bibinfo  {journal} {Ann. Rev. Astron. Astrophys.}\ }\textbf {\bibinfo {volume} {48}},\ \bibinfo {pages} {87} (\bibinfo {year} {2010})},\ \Eprint {https://arxiv.org/abs/1003.5567} {arXiv:1003.5567 [astro-ph.CO]} \BibitemShut {NoStop}%
\bibitem [{\citenamefont {Navarro}\ \emph {et~al.}(1996)\citenamefont {Navarro}, \citenamefont {Frenk},\ and\ \citenamefont {White}}]{Navarro:1995iw}%
  \BibitemOpen
  \bibfield  {author} {\bibinfo {author} {\bibfnamefont {J.~F.}\ \bibnamefont {Navarro}}, \bibinfo {author} {\bibfnamefont {C.~S.}\ \bibnamefont {Frenk}},\ and\ \bibinfo {author} {\bibfnamefont {S.~D.~M.}\ \bibnamefont {White}},\ }\bibfield  {title} {\bibinfo {title} {{The Structure of cold dark matter halos}},\ }\href {https://doi.org/10.1086/177173} {\bibfield  {journal} {\bibinfo  {journal} {Astrophys. J.}\ }\textbf {\bibinfo {volume} {462}},\ \bibinfo {pages} {563} (\bibinfo {year} {1996})},\ \Eprint {https://arxiv.org/abs/astro-ph/9508025} {arXiv:astro-ph/9508025} \BibitemShut {NoStop}%
\bibitem [{\citenamefont {Navarro}\ \emph {et~al.}(1997)\citenamefont {Navarro}, \citenamefont {Frenk},\ and\ \citenamefont {White}}]{Navarro:1996gj}%
  \BibitemOpen
  \bibfield  {author} {\bibinfo {author} {\bibfnamefont {J.~F.}\ \bibnamefont {Navarro}}, \bibinfo {author} {\bibfnamefont {C.~S.}\ \bibnamefont {Frenk}},\ and\ \bibinfo {author} {\bibfnamefont {S.~D.~M.}\ \bibnamefont {White}},\ }\bibfield  {title} {\bibinfo {title} {{A Universal density profile from hierarchical clustering}},\ }\href {https://doi.org/10.1086/304888} {\bibfield  {journal} {\bibinfo  {journal} {Astrophys. J.}\ }\textbf {\bibinfo {volume} {490}},\ \bibinfo {pages} {493} (\bibinfo {year} {1997})},\ \Eprint {https://arxiv.org/abs/astro-ph/9611107} {arXiv:astro-ph/9611107} \BibitemShut {NoStop}%
\bibitem [{\citenamefont {Villarrubia-Rojo}\ \emph {et~al.}(2024)\citenamefont {Villarrubia-Rojo}, \citenamefont {Savastano}, \citenamefont {Zumalac\'arregui}, \citenamefont {Choi}, \citenamefont {Goyal}, \citenamefont {Dai},\ and\ \citenamefont {Tambalo}}]{Villarrubia-Rojo:2024xcj}%
  \BibitemOpen
  \bibfield  {author} {\bibinfo {author} {\bibfnamefont {H.}~\bibnamefont {Villarrubia-Rojo}}, \bibinfo {author} {\bibfnamefont {S.}~\bibnamefont {Savastano}}, \bibinfo {author} {\bibfnamefont {M.}~\bibnamefont {Zumalac\'arregui}}, \bibinfo {author} {\bibfnamefont {L.}~\bibnamefont {Choi}}, \bibinfo {author} {\bibfnamefont {S.}~\bibnamefont {Goyal}}, \bibinfo {author} {\bibfnamefont {L.}~\bibnamefont {Dai}},\ and\ \bibinfo {author} {\bibfnamefont {G.}~\bibnamefont {Tambalo}},\ }\bibfield  {title} {\bibinfo {title} {{GLoW: novel methods for wave-optics phenomena in gravitational lensing}}\ }\href@noop {} {} (\bibinfo {year} {2024}),\ \Eprint {https://arxiv.org/abs/2409.04606} {arXiv:2409.04606 [gr-qc]} \BibitemShut {NoStop}%
\bibitem [{\citenamefont {Branchesi}\ \emph {et~al.}(2023)\citenamefont {Branchesi} \emph {et~al.}}]{Branchesi:2023mws}%
  \BibitemOpen
  \bibfield  {author} {\bibinfo {author} {\bibfnamefont {M.}~\bibnamefont {Branchesi}} \emph {et~al.},\ }\bibfield  {title} {\bibinfo {title} {{Science with the Einstein Telescope: a comparison of different designs}},\ }\href {https://doi.org/10.1088/1475-7516/2023/07/068} {\bibfield  {journal} {\bibinfo  {journal} {JCAP}\ }\textbf {\bibinfo {volume} {07}},\ \bibinfo {pages} {068}},\ \Eprint {https://arxiv.org/abs/2303.15923} {arXiv:2303.15923 [gr-qc]} \BibitemShut {NoStop}%
\bibitem [{\citenamefont {Hild}\ \emph {et~al.}(2011)\citenamefont {Hild} \emph {et~al.}}]{Hild:2010id}%
  \BibitemOpen
  \bibfield  {author} {\bibinfo {author} {\bibfnamefont {S.}~\bibnamefont {Hild}} \emph {et~al.},\ }\bibfield  {title} {\bibinfo {title} {{Sensitivity Studies for Third-Generation Gravitational Wave Observatories}},\ }\href {https://doi.org/10.1088/0264-9381/28/9/094013} {\bibfield  {journal} {\bibinfo  {journal} {Class. Quant. Grav.}\ }\textbf {\bibinfo {volume} {28}},\ \bibinfo {pages} {094013} (\bibinfo {year} {2011})},\ \Eprint {https://arxiv.org/abs/1012.0908} {arXiv:1012.0908 [gr-qc]} \BibitemShut {NoStop}%
\bibitem [{\citenamefont {Husa}\ \emph {et~al.}(2016)\citenamefont {Husa}, \citenamefont {Khan}, \citenamefont {Hannam}, \citenamefont {P\"urrer}, \citenamefont {Ohme}, \citenamefont {Jim\'enez~Forteza},\ and\ \citenamefont {Boh\'e}}]{Husa:2015iqa}%
  \BibitemOpen
  \bibfield  {author} {\bibinfo {author} {\bibfnamefont {S.}~\bibnamefont {Husa}}, \bibinfo {author} {\bibfnamefont {S.}~\bibnamefont {Khan}}, \bibinfo {author} {\bibfnamefont {M.}~\bibnamefont {Hannam}}, \bibinfo {author} {\bibfnamefont {M.}~\bibnamefont {P\"urrer}}, \bibinfo {author} {\bibfnamefont {F.}~\bibnamefont {Ohme}}, \bibinfo {author} {\bibfnamefont {X.}~\bibnamefont {Jim\'enez~Forteza}},\ and\ \bibinfo {author} {\bibfnamefont {A.}~\bibnamefont {Boh\'e}},\ }\bibfield  {title} {\bibinfo {title} {{Frequency-domain gravitational waves from nonprecessing black-hole binaries. I. New numerical waveforms and anatomy of the signal}},\ }\href {https://doi.org/10.1103/PhysRevD.93.044006} {\bibfield  {journal} {\bibinfo  {journal} {Phys. Rev. D}\ }\textbf {\bibinfo {volume} {93}},\ \bibinfo {pages} {044006} (\bibinfo {year} {2016})},\ \Eprint {https://arxiv.org/abs/1508.07250} {arXiv:1508.07250 [gr-qc]} \BibitemShut {NoStop}%
\bibitem [{\citenamefont {Khan}\ \emph {et~al.}(2016)\citenamefont {Khan}, \citenamefont {Husa}, \citenamefont {Hannam}, \citenamefont {Ohme}, \citenamefont {P\"urrer}, \citenamefont {Jim\'enez~Forteza},\ and\ \citenamefont {Boh\'e}}]{Khan:2015jqa}%
  \BibitemOpen
  \bibfield  {author} {\bibinfo {author} {\bibfnamefont {S.}~\bibnamefont {Khan}}, \bibinfo {author} {\bibfnamefont {S.}~\bibnamefont {Husa}}, \bibinfo {author} {\bibfnamefont {M.}~\bibnamefont {Hannam}}, \bibinfo {author} {\bibfnamefont {F.}~\bibnamefont {Ohme}}, \bibinfo {author} {\bibfnamefont {M.}~\bibnamefont {P\"urrer}}, \bibinfo {author} {\bibfnamefont {X.}~\bibnamefont {Jim\'enez~Forteza}},\ and\ \bibinfo {author} {\bibfnamefont {A.}~\bibnamefont {Boh\'e}},\ }\bibfield  {title} {\bibinfo {title} {{Frequency-domain gravitational waves from nonprecessing black-hole binaries. II. A phenomenological model for the advanced detector era}},\ }\href {https://doi.org/10.1103/PhysRevD.93.044007} {\bibfield  {journal} {\bibinfo  {journal} {Phys. Rev. D}\ }\textbf {\bibinfo {volume} {93}},\ \bibinfo {pages} {044007} (\bibinfo {year} {2016})},\ \Eprint {https://arxiv.org/abs/1508.07253} {arXiv:1508.07253 [gr-qc]} \BibitemShut {NoStop}%
\bibitem [{\citenamefont {Nitz}\ \emph {et~al.}(2024)\citenamefont {Nitz}, \citenamefont {Harry}, \citenamefont {Brown}, \citenamefont {Biwer}, \citenamefont {Willis}, \citenamefont {Canton}, \citenamefont {Capano}, \citenamefont {Dent}, \citenamefont {Pekowsky}, \citenamefont {Davies}, \citenamefont {De}, \citenamefont {Cabero}, \citenamefont {Wu}, \citenamefont {Williamson}, \citenamefont {Machenschalk}, \citenamefont {Macleod}, \citenamefont {Pannarale}, \citenamefont {Kumar}, \citenamefont {Reyes},\ and\ \citenamefont {Tolley}}]{nitz2024pycbc}%
  \BibitemOpen
  \bibfield  {author} {\bibinfo {author} {\bibfnamefont {A.}~\bibnamefont {Nitz}}, \bibinfo {author} {\bibfnamefont {I.}~\bibnamefont {Harry}}, \bibinfo {author} {\bibfnamefont {D.}~\bibnamefont {Brown}}, \bibinfo {author} {\bibfnamefont {C.~M.}\ \bibnamefont {Biwer}}, \bibinfo {author} {\bibfnamefont {J.}~\bibnamefont {Willis}}, \bibinfo {author} {\bibfnamefont {T.~D.}\ \bibnamefont {Canton}}, \bibinfo {author} {\bibfnamefont {C.}~\bibnamefont {Capano}}, \bibinfo {author} {\bibfnamefont {T.}~\bibnamefont {Dent}}, \bibinfo {author} {\bibfnamefont {L.}~\bibnamefont {Pekowsky}}, \bibinfo {author} {\bibfnamefont {G.~S.~C.}\ \bibnamefont {Davies}}, \bibinfo {author} {\bibfnamefont {S.}~\bibnamefont {De}}, \bibinfo {author} {\bibfnamefont {M.}~\bibnamefont {Cabero}}, \bibinfo {author} {\bibfnamefont {S.}~\bibnamefont {Wu}}, \bibinfo {author} {\bibfnamefont {A.~R.}\ \bibnamefont {Williamson}}, \bibinfo {author} {\bibfnamefont {B.}~\bibnamefont {Machenschalk}}, \bibinfo {author} {\bibfnamefont {D.}~\bibnamefont
  {Macleod}}, \bibinfo {author} {\bibfnamefont {F.}~\bibnamefont {Pannarale}}, \bibinfo {author} {\bibfnamefont {P.}~\bibnamefont {Kumar}}, \bibinfo {author} {\bibfnamefont {S.}~\bibnamefont {Reyes}},\ and\ \bibinfo {author} {\bibfnamefont {A.}~\bibnamefont {Tolley}},\ }\href@noop {} {\bibinfo {title} {gwastro/pycbc: v2.3.3 release of pycbc (v2.3.3)}},\ \bibinfo {howpublished} {\url{https://doi.org/10.5281/zenodo.10473621}} (\bibinfo {year} {2024}),\ \bibinfo {note} {zenodo}\BibitemShut {NoStop}%
\bibitem [{\citenamefont {Nakamura}\ \emph {et~al.}(2016)\citenamefont {Nakamura} \emph {et~al.}}]{Nakamura:2016hna}%
  \BibitemOpen
  \bibfield  {author} {\bibinfo {author} {\bibfnamefont {T.}~\bibnamefont {Nakamura}} \emph {et~al.},\ }\bibfield  {title} {\bibinfo {title} {{Pre-DECIGO can get the smoking gun to decide the astrophysical or cosmological origin of GW150914-like binary black holes}},\ }\href {https://doi.org/10.1093/ptep/ptw127} {\bibfield  {journal} {\bibinfo  {journal} {PTEP}\ }\textbf {\bibinfo {volume} {2016}},\ \bibinfo {pages} {093E01} (\bibinfo {year} {2016})},\ \Eprint {https://arxiv.org/abs/1607.00897} {arXiv:1607.00897 [astro-ph.HE]} \BibitemShut {NoStop}%
\bibitem [{\citenamefont {Yagi}\ and\ \citenamefont {Seto}(2011)}]{Yagi:2011wg}%
  \BibitemOpen
  \bibfield  {author} {\bibinfo {author} {\bibfnamefont {K.}~\bibnamefont {Yagi}}\ and\ \bibinfo {author} {\bibfnamefont {N.}~\bibnamefont {Seto}},\ }\bibfield  {title} {\bibinfo {title} {{Detector configuration of DECIGO/BBO and identification of cosmological neutron-star binaries}},\ }\href {https://doi.org/10.1103/PhysRevD.83.044011} {\bibfield  {journal} {\bibinfo  {journal} {Phys. Rev. D}\ }\textbf {\bibinfo {volume} {83}},\ \bibinfo {pages} {044011} (\bibinfo {year} {2011})},\ \bibinfo {note} {[Erratum: Phys.Rev.D 95, 109901 (2017)]},\ \Eprint {https://arxiv.org/abs/1101.3940} {arXiv:1101.3940 [astro-ph.CO]} \BibitemShut {NoStop}%
\bibitem [{\citenamefont {Vallisneri}(2008)}]{Vallisneri:2007ev}%
  \BibitemOpen
  \bibfield  {author} {\bibinfo {author} {\bibfnamefont {M.}~\bibnamefont {Vallisneri}},\ }\bibfield  {title} {\bibinfo {title} {{Use and abuse of the Fisher information matrix in the assessment of gravitational-wave parameter-estimation prospects}},\ }\href {https://doi.org/10.1103/PhysRevD.77.042001} {\bibfield  {journal} {\bibinfo  {journal} {Phys. Rev. D}\ }\textbf {\bibinfo {volume} {77}},\ \bibinfo {pages} {042001} (\bibinfo {year} {2008})},\ \Eprint {https://arxiv.org/abs/gr-qc/0703086} {arXiv:gr-qc/0703086} \BibitemShut {NoStop}%
\bibitem [{\citenamefont {Rodriguez}\ \emph {et~al.}(2013)\citenamefont {Rodriguez}, \citenamefont {Farr}, \citenamefont {Farr},\ and\ \citenamefont {Mandel}}]{Rodriguez:2013mla}%
  \BibitemOpen
  \bibfield  {author} {\bibinfo {author} {\bibfnamefont {C.~L.}\ \bibnamefont {Rodriguez}}, \bibinfo {author} {\bibfnamefont {B.}~\bibnamefont {Farr}}, \bibinfo {author} {\bibfnamefont {W.~M.}\ \bibnamefont {Farr}},\ and\ \bibinfo {author} {\bibfnamefont {I.}~\bibnamefont {Mandel}},\ }\bibfield  {title} {\bibinfo {title} {{Inadequacies of the Fisher Information Matrix in gravitational-wave parameter estimation}},\ }\href {https://doi.org/10.1103/PhysRevD.88.084013} {\bibfield  {journal} {\bibinfo  {journal} {Phys. Rev. D}\ }\textbf {\bibinfo {volume} {88}},\ \bibinfo {pages} {084013} (\bibinfo {year} {2013})},\ \Eprint {https://arxiv.org/abs/1308.1397} {arXiv:1308.1397 [astro-ph.IM]} \BibitemShut {NoStop}%
\bibitem [{\citenamefont {Cutler}\ and\ \citenamefont {Flanagan}(1994)}]{Cutler:1994ys}%
  \BibitemOpen
  \bibfield  {author} {\bibinfo {author} {\bibfnamefont {C.}~\bibnamefont {Cutler}}\ and\ \bibinfo {author} {\bibfnamefont {E.~E.}\ \bibnamefont {Flanagan}},\ }\bibfield  {title} {\bibinfo {title} {{Gravitational waves from merging compact binaries: How accurately can one extract the binary's parameters from the inspiral wave form?}},\ }\href {https://doi.org/10.1103/PhysRevD.49.2658} {\bibfield  {journal} {\bibinfo  {journal} {Phys. Rev. D}\ }\textbf {\bibinfo {volume} {49}},\ \bibinfo {pages} {2658} (\bibinfo {year} {1994})},\ \Eprint {https://arxiv.org/abs/gr-qc/9402014} {arXiv:gr-qc/9402014} \BibitemShut {NoStop}%
\bibitem [{\citenamefont {Lee}\ \emph {et~al.}(2022)\citenamefont {Lee}, \citenamefont {Morisaki},\ and\ \citenamefont {Tagoshi}}]{Lee:2022jpn}%
  \BibitemOpen
  \bibfield  {author} {\bibinfo {author} {\bibfnamefont {E.}~\bibnamefont {Lee}}, \bibinfo {author} {\bibfnamefont {S.}~\bibnamefont {Morisaki}},\ and\ \bibinfo {author} {\bibfnamefont {H.}~\bibnamefont {Tagoshi}},\ }\bibfield  {title} {\bibinfo {title} {{Mass-spin reparametrization for a rapid parameter estimation of inspiral gravitational-wave signals}},\ }\href {https://doi.org/10.1103/PhysRevD.105.124057} {\bibfield  {journal} {\bibinfo  {journal} {Phys. Rev. D}\ }\textbf {\bibinfo {volume} {105}},\ \bibinfo {pages} {124057} (\bibinfo {year} {2022})},\ \Eprint {https://arxiv.org/abs/2203.05216} {arXiv:2203.05216 [gr-qc]} \BibitemShut {NoStop}%
\end{thebibliography}%
\clearpage

\appendix
\onecolumngrid
\section*{Supplement materials}
\label{appendix: supplement}

\begin{figure}[h] 
	\centering
	\includegraphics[width=\hsize]{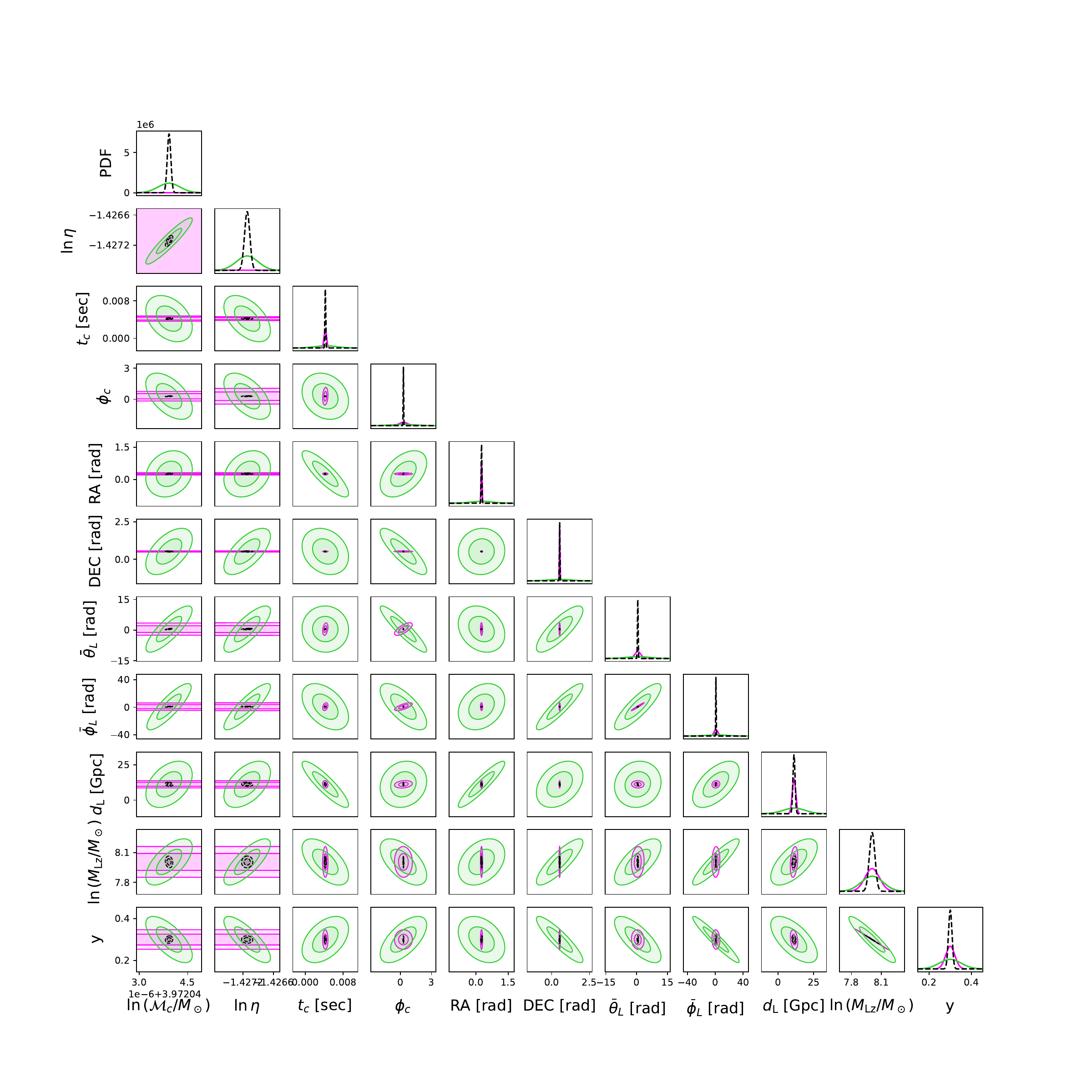}
	\caption{$50~\%$ and $90~\%$ error contours assuming the \ac{DECIGO} and  the \ac{ET} observations in SIS lens model. The magenta and the green contours show the Fisher forecasts for \ac{ET} and \ac{DECIGO} respectively. The black dashed contour represents the joint analysis of \ac{ET} and \ac{DECIGO}.}
	\label{fig: Fisher_SIS_ET_DECIGO_all}
\end{figure}
\clearpage

\begin{figure}[h] 
	\centering
	\includegraphics[width=\hsize]{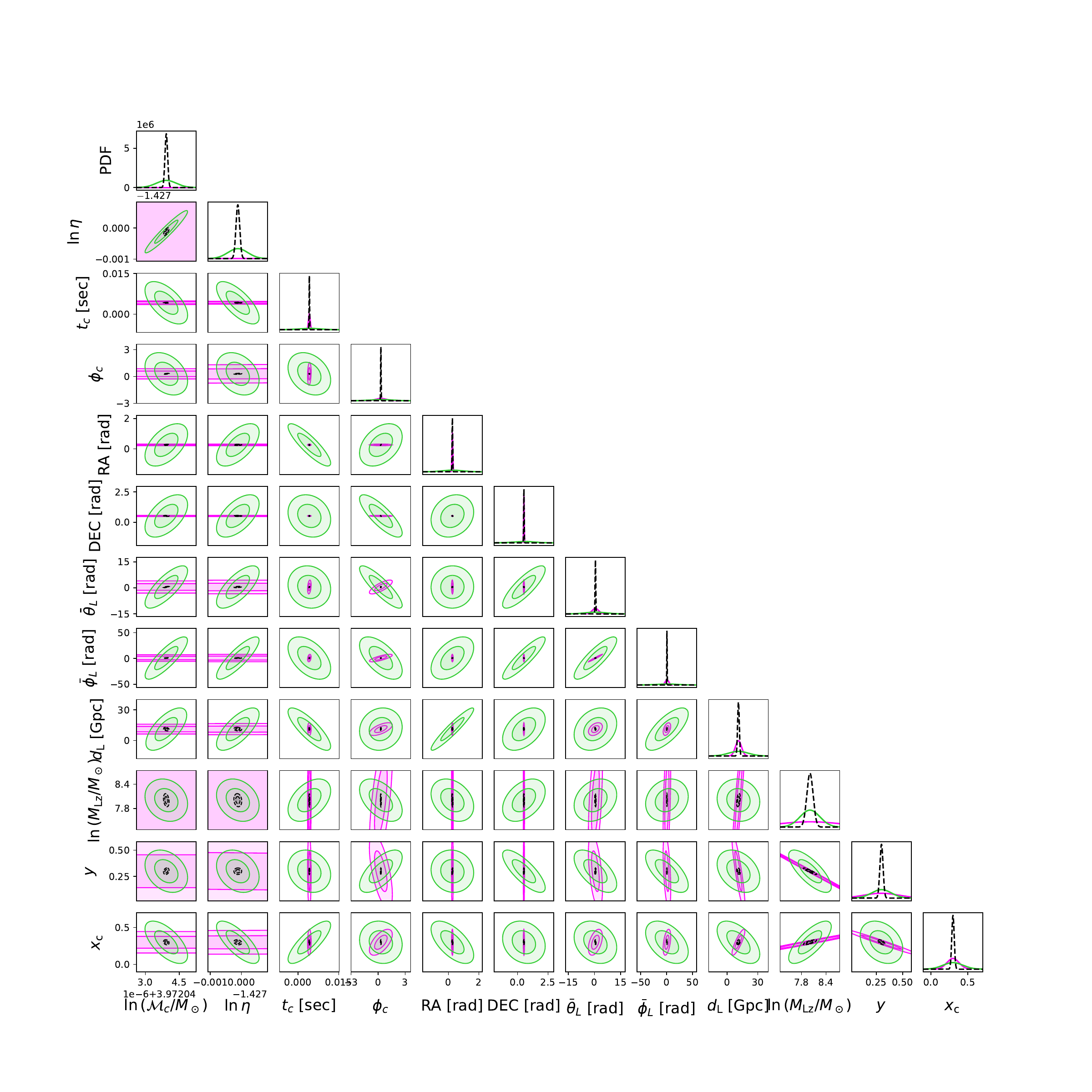}
	\caption{$50~\%$ and $90~\%$ error contours assuming the \ac{DECIGO} and  the \ac{ET} observations in CIS lens model. The magenta and the green contours show the Fisher forecasts for \ac{ET} and \ac{DECIGO}, respectively. The black dashed contour represents the joint analysis of \ac{ET} and \ac{DECIGO}.}
	\label{fig: Fisher_CIS_ET_DECIGO_all}
\end{figure}
\clearpage

\begin{figure}[h] 
	\centering
	\includegraphics[width=\hsize]{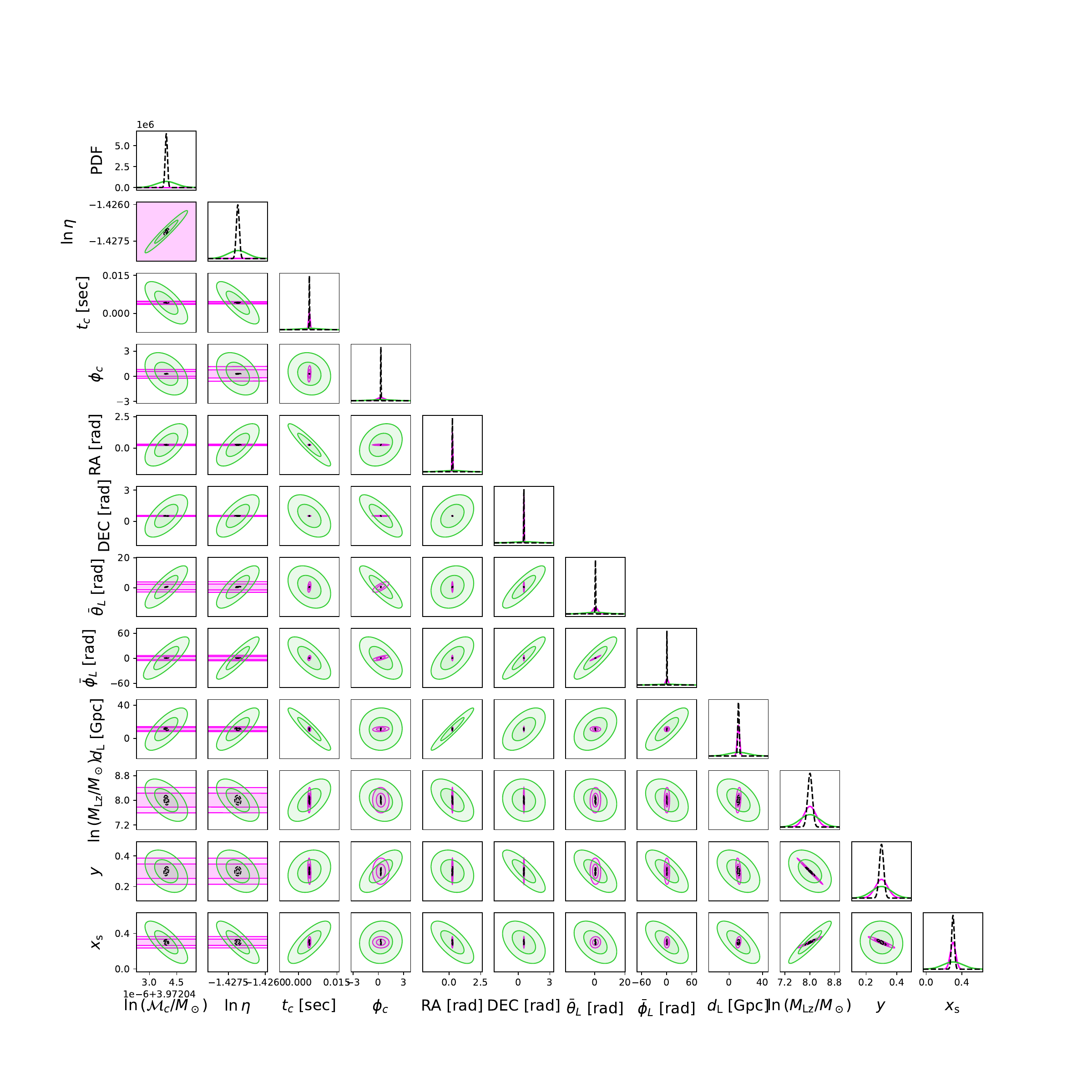}
	\caption{$50~\%$ and $90~\%$ error contours assuming the \ac{DECIGO} and  the \ac{ET} observations in NFW lens model. The magenta and the green contours show the Fisher forecasts for \ac{ET} and \ac{DECIGO} respectively. The black dashed contour represents the joint analysis of \ac{ET} and \ac{DECIGO}.}
	\label{fig: Fisher_NFW_ET_DECIGO_all}
\end{figure}
\clearpage

\begin{figure}[htbp]
    \centering
    \begin{tabular}{cc}
        % Row 1 (SIS model)
        \includegraphics[width=0.48\textwidth]{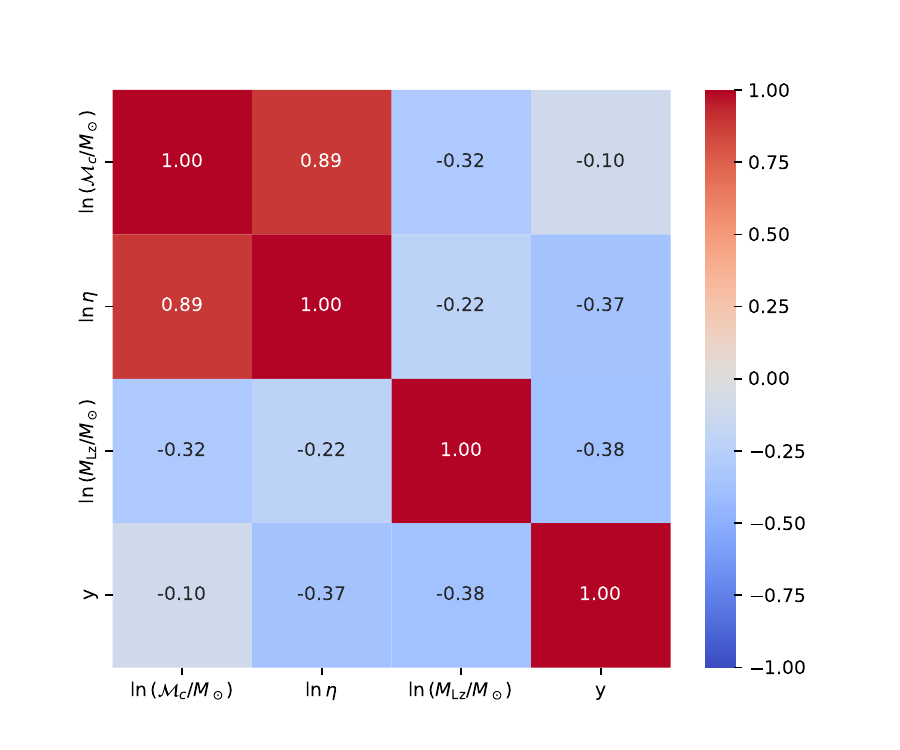} &
        \includegraphics[width=0.48\textwidth]{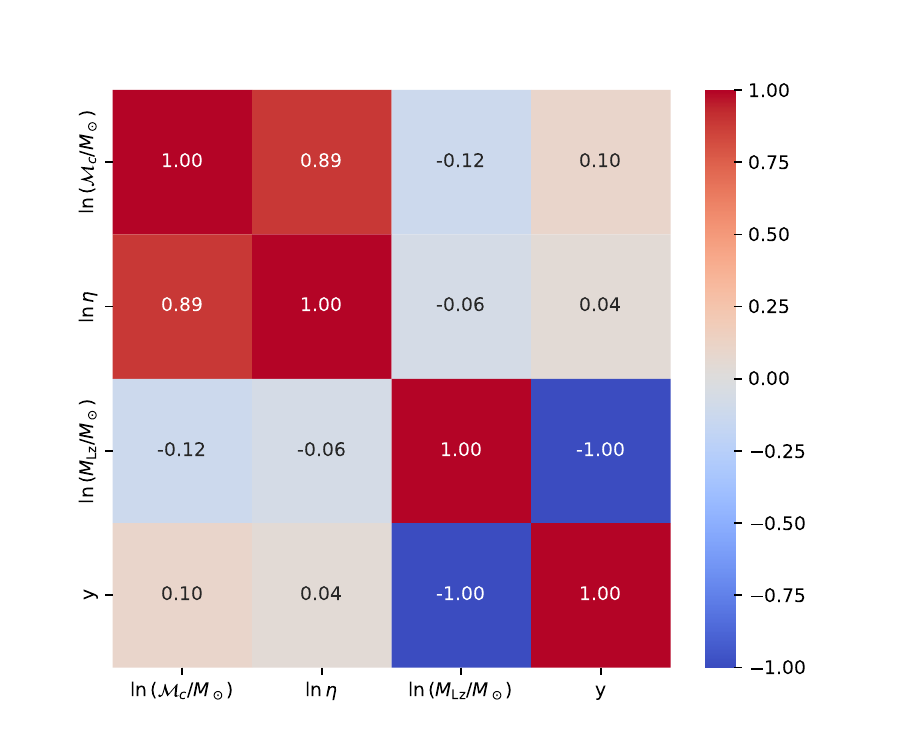} \\[1ex]
        % Row 2 (CIS model)
        \includegraphics[width=0.48\textwidth]{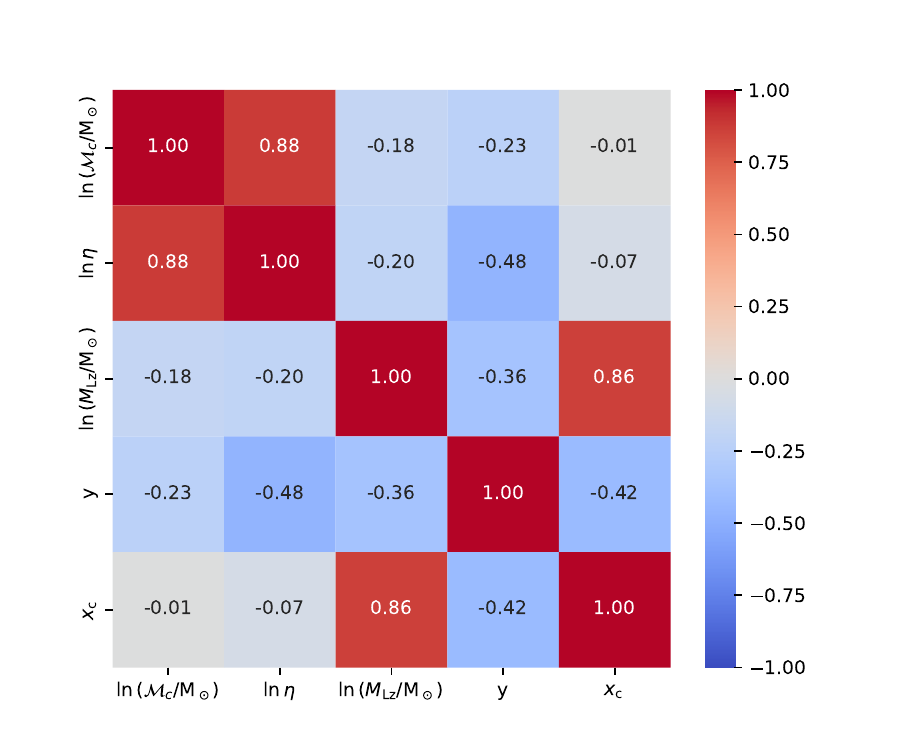} &
        \includegraphics[width=0.48\textwidth]{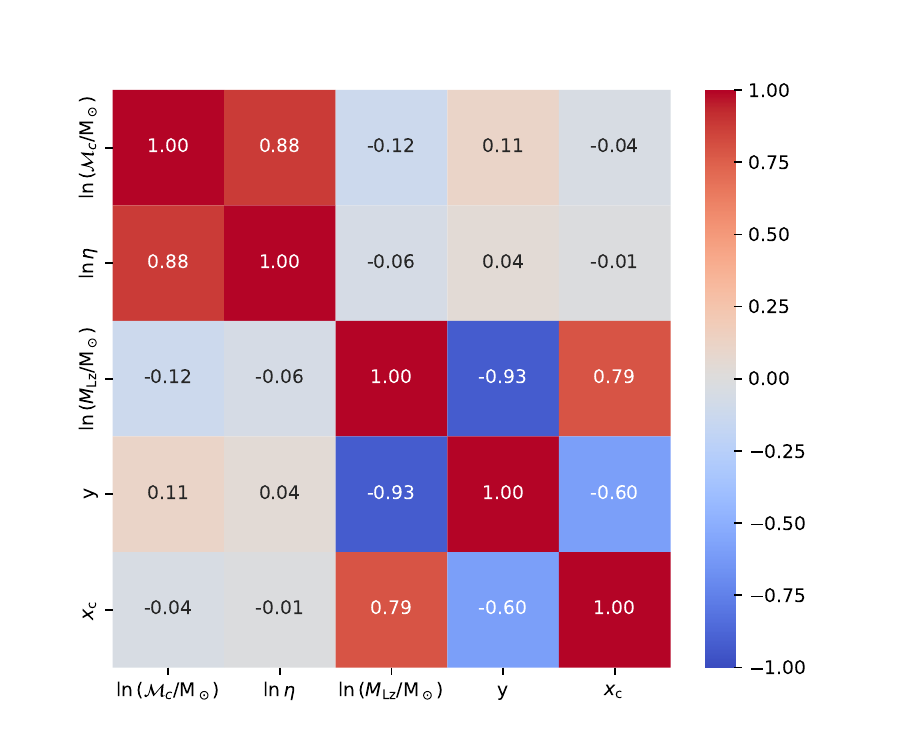} \\[1ex]
        % Row 3 (NFW model)
        \includegraphics[width=0.48\textwidth]{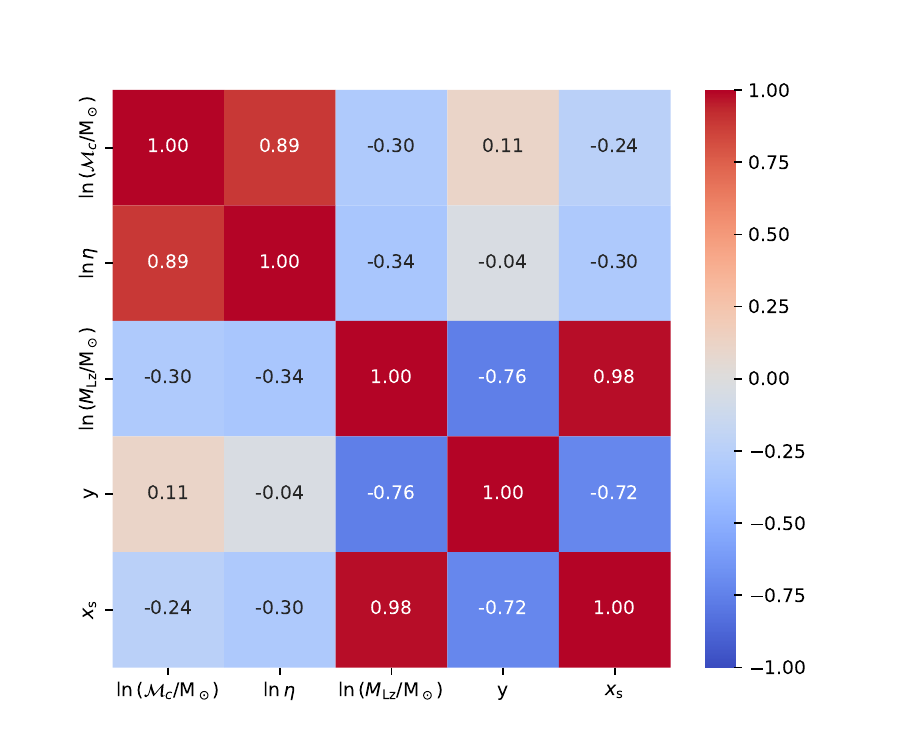} &
        \includegraphics[width=0.48\textwidth]{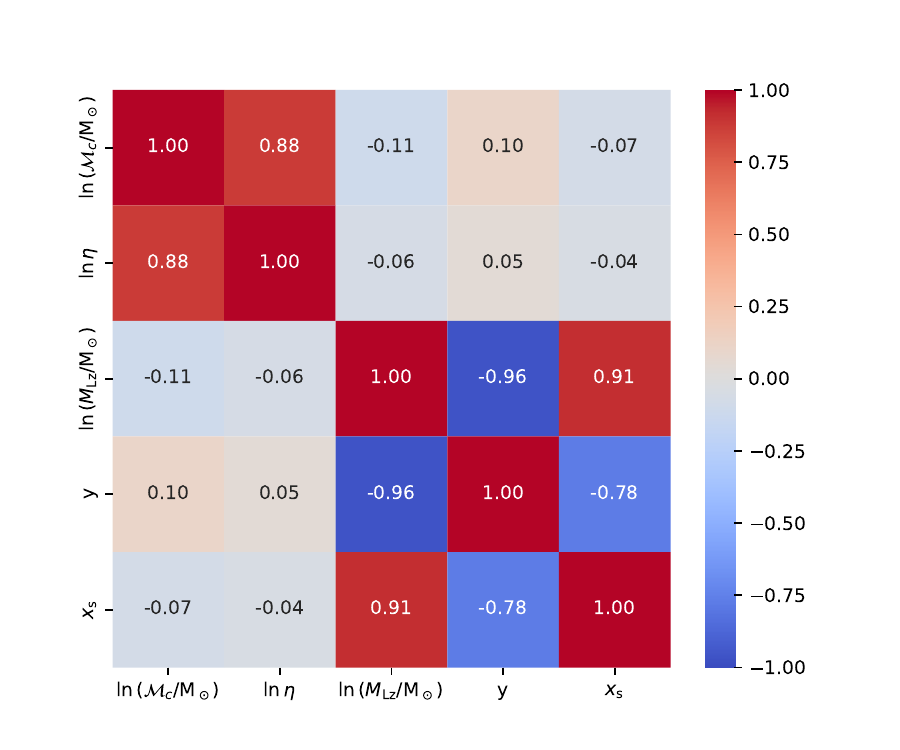} \\
    \end{tabular}
    \caption{%
    Correlation coefficients for each lens model. 
    \emph{Top row (SIS):} 
      \emph{Left}: B-\ac{DECIGO} alone; 
      \emph{Right}: B-\ac{DECIGO} + \ac{ET}. 
    \emph{Middle row (CIS):} 
      \emph{Left}: B-\ac{DECIGO} alone; 
      \emph{Right}: B-\ac{DECIGO} + \ac{ET}. 
    \emph{Bottom row (NFW):} 
      \emph{Left}: B-\ac{DECIGO} alone; 
      \emph{Right}: B-\ac{DECIGO} + \ac{ET}.}
    \label{fig:corr_all_models_BDECIGO}
\end{figure}

%%ET vs. DECIGO
\begin{figure}[htbp]
    \centering
    \begin{tabular}{cc}
        
        \includegraphics[width=0.48\textwidth]{Fig/corr_BDECIGO_SIS.pdf} &
        \includegraphics[width=0.48\textwidth]{Fig/corr_ET_BDECIGO_SIS.pdf} \\[1ex]
        % Row 2 (CIS model)
        \includegraphics[width=0.48\textwidth]{Fig/corr_BDECIGO_CIS.pdf} &
        \includegraphics[width=0.48\textwidth]{Fig/corr_ET_BDECIGO_CIS.pdf} \\[1ex]
        % Row 3 (NFW model)
        \includegraphics[width=0.48\textwidth]{Fig/corr_BDECIGO_NFW.pdf} &
        \includegraphics[width=0.48\textwidth]{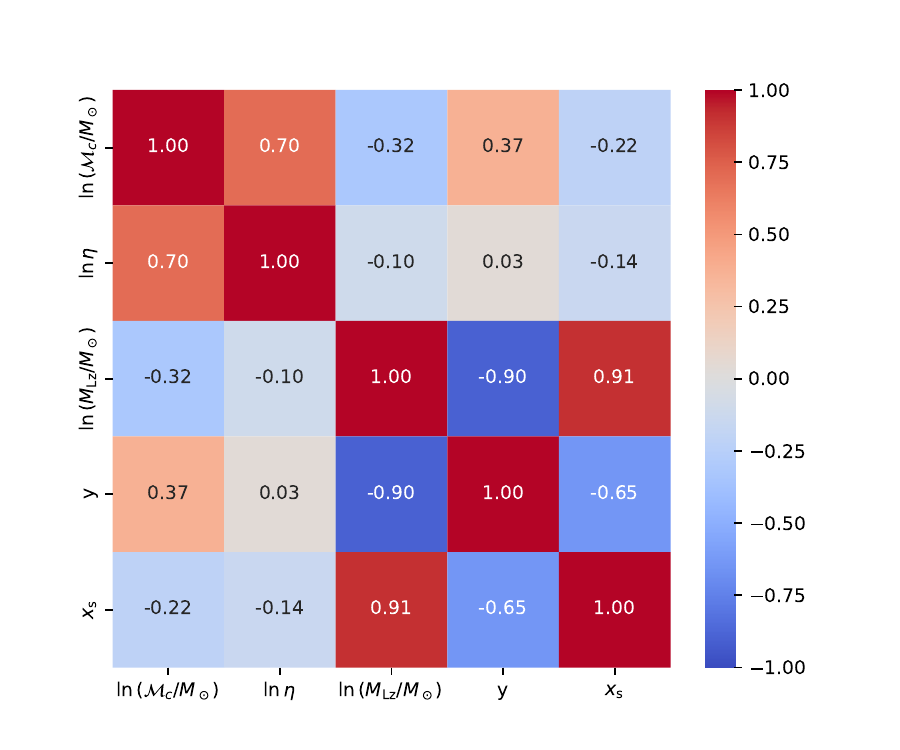} \\
    \end{tabular}
    \caption{%
    Correlation coefficients for each lens model. 
    \emph{Top row (SIS):} 
      \emph{Left}: \ac{DECIGO} alone; 
      \emph{Right}: \ac{DECIGO} + \ac{ET}. 
    \emph{Middle row (CIS):} 
      \emph{Left}: \ac{DECIGO} alone; 
      \emph{Right}: \ac{DECIGO} + \ac{ET}. 
    \emph{Bottom row (NFW):} 
      \emph{Left}: \ac{DECIGO} alone; 
      \emph{Right}: \ac{DECIGO} + \ac{ET}.}
    \label{fig:corr_all_models_DECIGO}
\end{figure}
\clearpage

\end{document}